%% file: main.tex
\documentclass{IEEEaccess}
\usepackage{cite}
\usepackage{amsmath,amssymb,amsfonts}
\usepackage{algorithmic}
\usepackage{graphicx}
\usepackage{textcomp}
\usepackage{braket}
\usepackage{bm}
\usepackage{listings}
\usepackage{hyperref}

\def\BibTeX{{\rm B\kern-.05em{\sc i\kern-.025em b}\kern-.08em
    T\kern-.1667em\lower.7ex\hbox{E}\kern-.125emX}}
\begin{document}
%\history{Date of publication xxxx 00, 0000, date of current version xxxx 00, 0000.}
%\history{}
%\doi{10.1109/TQE.2020.DOI}

\title{QPack: Quantum Approximate Optimization Algorithms as benchmark for quantum computers}
\author{\uppercase{Koen Mesman}\authorrefmark{1},
\uppercase{Huub Donkers\authorrefmark{1}, Zaid Al-Ars\authorrefmark{1} and Matthias Möller}.\authorrefmark{2}}
\address[1]{Delft University of Technology, department of Computer Engineering}
\address[2]{Delft University of Technology, department of Applied Mathematics}

\corresp{Corresponding author: Koen Mesman (email: kmesman@tudelft.nl).}

\markboth
{Mesman \headeretal: Preparation of Papers for IEEE Transactions on Quantum Engineering}
{Mesman \headeretal: Preparation of Papers for IEEE Transactions on Quantum Engineering}

\begin{abstract}
%This report aims to give the proper preliminary information on Quantum Approximation Optimization Algorithms (QAOA) and explore the applications and performance of these applications. The state of the art implementations are evaluated according to their technology readiness level to give insight on the achievements and  limitations of current QAOA implementations. The current challenges are evaluated and emerging techniques to improve application are presented, such as machine learning for parameter optimization. Currently, the largest obstacles in QAOA are the optimization of parameter $p$, determining the QAOA cycles on the quantum computer, and the limitations of current quantum hardware.
\noindent In this paper, we present QPack, a universal benchmark for Noisy Intermediate-Scale Quantum (NISQ) computers based on Quantum Approximate Optimization Algorithms (QAOA). Unlike other evaluation metrics in the field, this benchmark evaluates not only one, but multiple important aspects of quantum computing hardware: the maximum problem size a quantum computer can solve, the required runtime, as well as the achieved accuracy. The applications MaxCut, dominating set and traveling salesman are included to provide variation in resource requirements. This will allow for a diverse benchmark that promotes optimal design considerations, avoiding hardware implementations for specific applications. We also discuss the design aspects that are taken in consideration for the QPack benchmark, with critical quantum benchmark requirements in mind. An implementation is presented, providing practical metrics.  QPack is presented as a hardware agnostic benchmark by making use of the XACC library. We demonstrate the application of the benchmark on various IBM machines, as well as a range of simulators.
\end{abstract}


\titlepgskip=-15pt

\maketitle

\input{Intro_benchmark}
\input{Selecting_QA}

\input{QTRL}
\input{benchmark_updated}
\input{results}
\input{Conclusion}

\bibliographystyle{IEEEtran}
\bibliography{bib}

\begin{appendices}
\newpage
\input{QAOA_theory}
\input{QAOA_examples}
\input{Selecting_app}

\input{Parameter_optimization}
%\input{code_appendix}
\input{results_appendix}

\input{additional_XACC_results}
\end{appendices}

\EOD

\end{document}

%% file: Intro_benchmark.tex
\section{Towards a standard quantum benchmark}
\label{sec:intro}
%Current quantum benchmarks & concerns

\noindent Currently, quantum computing is making large steps to becoming a mature technology. Many companies are developing their own quantum hardware for both research and preparing for practical deployment. The main challenges in the past years have been an abundance of practical applications for the few-qubit Noisy Intermediate-Scale Quantum (NISQ) quantum hardware and scaling the qubits and depth of the quantum hardware. For these reasons, quantum computing was not yet at the maturity level needed to create a practical application level quantum benchmark, i.e. a benchmark that represents practical utility.\\

Many quantum benchmarks have been developed, but these either aim at qubit level assessment or are lacking in practical usability. Benchmarks at the component level (individual qubits, quantum logic gates) have been developed widely and are useful for the development of the hardware. This is, however, more aimed at quantum research on low-level hardware rather than application \cite{blume2019metrics}. This also serves a fundamentally different goal than a performance measure of quantum computers. Many different forms of quantum hardware are being used, without a clear dominating implementation. Each of these have different dynamics and metrics, and can therefore not be generalized \cite{resch2019benchmarking}. 
%For this reason, making a benchmark for low-level hardware aspects will not properly reflect its performance compared to other implementations of the quantum hardware.
%Toegevoegd:
\iffalse
Determining the performance of the complete system will therefore require abstraction from the low level performance and an application level benchmark is needed to properly verify the performance.\\
\fi
%\iffalse %Misschien een beetje dubbel, stukje toegevoegd aan vorige alinea
Furthermore, while benchmarks for single qubit operations have been developed \cite{rudinger2016quantum, knill2008randomized, dankert2009exact, emerson2005scalable}, these performances do not accurately reflect quantum operations on larger scales. Noise levels, for example, are an important metric in single-gate performance. The noise, however, varies per gate and due to qubit entanglement will propagate unpredictably throughout the system \cite{resch2019benchmarking, erhard2019characterizing, ferracin2019accrediting}. This makes it impossible to use single gate noise performance to extrapolate for the entire system, while for classical computers this would have been possible. These difficulties in determining the performance will require abstraction from the low level performance and an application level benchmark is needed to properly verify the performance. \\
%\fi

%%%%%%%%%%%%%
%%%%%%%%%%%%%
%Zaid
%Please add a paragraph on the requirements that we need for a standardized benchmark. Then make a figure with a block diagram showing the different components needed in the benchmark, and identify in which section these blocks are discussed.
%%%%%%%%%%%%%%%%
%%%%%%%%%%%%%%%%

%\begin{figure}[h]
%    \centering
%    \includegraphics[clip, trim=1cm 11cm 1cm 0.5cm, width=0.7\linewidth]{figures/Benchmark_schematic.pdf}
%    \caption{Schematic outline of the QPack benchmark}
%    \label{fig:QPack_schematic}
%\end{figure}

\Figure[t!](topskip=0pt, botskip=0pt, midskip=0pt)[clip, trim=1cm 11cm 1cm 0.5cm, width=0.7\linewidth]{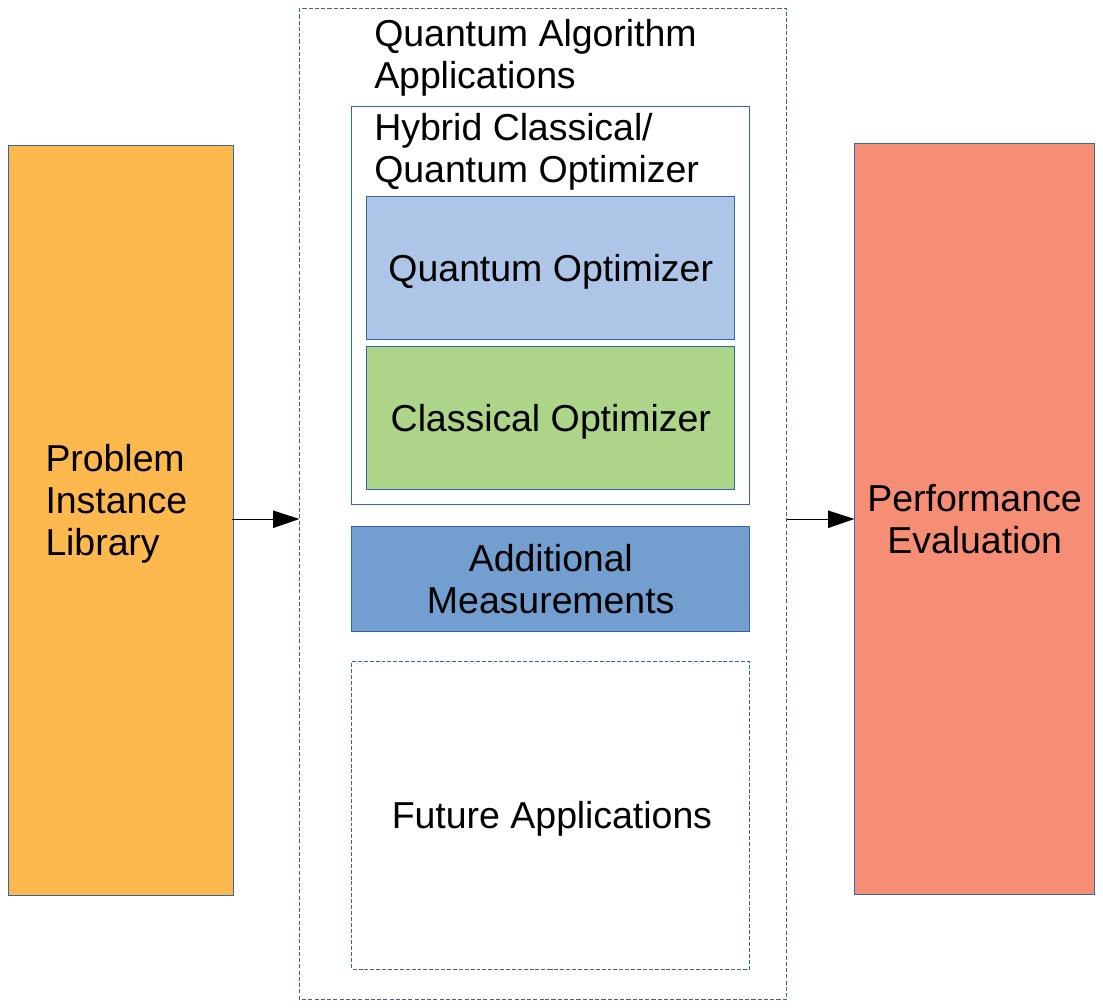}
{Schematic outline of the QPack benchmark\label{fig:QPack_schematic}}

In this paper, we propose QPack: a benchmark targeted to measure the performance of applications executed on quantum computers and quantum computer simulators. The benchmark has three main components, as presented in Figure~\ref{fig:QPack_schematic}: The problem library, the Quantum Algorithm and the Performance algorithm. The problem library contains a set of problems to be evaluated by the quantum algorithm. Alternative problems targeted by our benchmark are discussed in Section \ref{sec:QTRL} and Appendix \ref{sec:selecting_app}. The selected quantum optimizer is QAOA (Quantum Approximate Optimization Algorithm) \cite{farhi2014quantum}. The selection of this optimizer is further elaborated in Section \ref{sec:Quantum_Alg}. The performance evaluation of QAOA on a quantum device or simulator needs to be evaluated according to a selection of metrics. Further discussion on these metrics is presented in Section \ref{sec:lit_rev} and \ref{metrics}.\\

\iffalse
\noindent \textbf{Problem library}: A set of problems are provided to ensure practical relevance of the benchmark and a diverse means of evaluation. Alternative problems targeted by our benchmark are discussed in Section \ref{sec:QTRL} and Appendix \ref{sec:selecting_app}.\\

\noindent \textbf{Quantum algorithm}: A quantum algorithm must be chosen which can be applied to various problems and is applicable on current quantum hardware and quantum simulators. With current quantum hardware in mind, a hybrid classical-quantum optimizer is chosen as our first quantum algorithm. The selected quantum optimizer is QAOA (Quantum Approximate Optimization Algorithm) \cite{farhi2014quantum}. The selection of this optimizer is further elaborated in Section \ref{sec:Quantum_Alg} and \ref{sec:details_qaoa}. The options and selection of the classical optimizer are discussed in Appendix \ref{sec:param_opt}. Measurements of these classical optimizers are done as well, which is further elaborated in Section \ref{sec:measurements}. The QPack benchmark is envisioned to grow as more quantum algorithms become applicable on quantum hardware.\\

\noindent \textbf{Performance evaluation}: The performance of quantum computers and simulators need to be evaluated according to a selection of metrics. Further discussion on these metrics is presented in Section \ref{sec:lit_rev} and \ref{metrics}.\\
\fi

In this paper, a set of metrics is proposed to suit the emerging quantum QAOA approach. QAOA is a promising quantum approach for optimization using relatively few qubits. With the current rate of development of scaling quantum hardware, this approach is expected to be one of the first to have practical applications. The contributions of this work are as follows.
\begin{itemize}
    \item The general outline of the benchmark QPack, targeting essential metrics
    \item Analysis of classical optimizers to improve QAOA implementations
    \item Introducing the concept of quantum algorithm technology readiness level (QTRL), which gives a ranking of cutting edge QAOA implementations
    \item Demonstration of the hardware agnostic benchmark, with results on IBM hardware and various simulators.
\end{itemize}

QAOA algorithms are implemented in a hybrid classical-quantum computing set-up, where a problem is defined in the classical computer, after which an approximate optimization is performed using QAOA, of which the parameters are optimized using the classical computer. Generally, the classical and quantum computer work in an alternating fashion until an optimal configuration is found. QAOA has a limited probability of finding the optimal solution, but aims to achieve a equal or higher probability than its classical counterpart with improved performance. When the QAOA algorithm is executed, it can be assessed, used and reinitialized from the classical computer.  To further assess QAOA, the practical concerns will be covered as well. This will cover issues such as parameter tuning, but also concerns with current quantum hardware and how to deal with its limitations. In current trends, not only are there newer implementations, but also various ways to improve on the original algorithm. The shortcomings will be discussed and emerging trends such as QUBO \cite{Marshall_2020}, Quantum Alternating Operator Ansatz \cite{Hadfield_2019} and machine learning \cite{alam2020accelerating, Khairy2019ReinforcementLF,Wauters_2020} to improve upon the shortcomings are discussed.\\

In Section \ref{sec:lit_rev}, a literature review on quantum benchmarks is presented, addressing concerns on current quantum benchmarks and providing the necessary metrics.
Near term quantum algorithms will be discussed in Section \ref{sec:Quantum_Alg} including details on QAOA. 
In Section \ref{sec:QTRL}, the readiness of the implementations of the QAOA algorithm are evaluated, presenting a Quantum Technology Readiness Level (QTRL). This will give insight to how far the development of QAOA algorithms is, and whether these will be implemented in the near future.
The critical metrics and methods to reliably test these, are addressed in Section \ref{metrics}.
In Section \ref{sec:measurements}, classical optimizers and QAOA applications are respectively measured. From these measurements, insights are given for efficient implementations and possible further improvements.
The current QPack implementation is presented in Section \ref{sec:implementation}. The results of the benchmarks are presented in Section \ref{sec:results}, where the results on various IBM hardware are presented, as well as on various simulators.In Section \ref{sec:conclusion_future}, the conclusions and future work regarding the benchmark are discussed. \\

\section{Review of current quantum benchmarking} \label{sec:lit_rev}

Other benchmarks such as Random Benchmarking (RB) or quantum volume \cite{blume2020volumetric} are well known quantum benchmarks, but provide limited insight into practical use of quantum computers. RB is an example of a volumetric benchmark, which determines to which circuit depth the quantum hardware can hold the required fidelity for a set number of qubits. RB can determine the fidelity for a random sequence of quantum gates, which should give a general idea of how the hardware performs. By using RB, or any other volumetric benchmark, the quantum volume can be determined. However, this gives little insight on the performance of a practical algorithm on the quantum hardware.
Another challenge in developing a benchmark for quantum computing, is that there is no consensus on which metrics should be benchmarked to reflect performance properly.\\

\noindent As mentioned before, a popular metric to measure the performance of quantum computers is its quantum volume: the number of qubits and the quantum circuit depth it can run. The authors of the quantum volume benchmark mention in total four metrics to which the performance of a quantum computer can be evaluated \cite{bishop2017quantum}:
\begin{itemize}
    \item Number of physical qubits
    \item Number of gates that can be applied before errors occur (quantum circuit depth)
    \item Connectivity of the quantum computer
    \item Number of operations that can be run in parallel
\end{itemize}

The number of physical qubits is only of interest if a required circuit depth, as such these metrics can be combined in the quantum volume. As argued before, quantum volume on itself does not reflect practical performance properly. Furthermore, the qubit connectivity and number of operations that can be run in parallel certainly have value as metrics, however, it can again be argued whether these metrics have practical value. For example, a fully connected quantum computer is good on paper, but if it cannot scale up it cannot be put in practice. From the perspective of this paper, it is considered that execution with limited connectivity will be handled by the compiler (additional qubit SWAPs to enable execution on limited connectivity). As a result, the connectivity will affect the number of quantum operations and with that the run time and outcome accuracy (increasing SWAP-gates and in turn circuit depth will give rise to more noise, decreasing outcome accuracy). An important characteristic of QPack, is that it will use non-native problems, meaning that the problem does not correspond to the qubit layout. This means more resources are required to compile the problem to the quantum computer's hardware layout, which significantly impacts performance \cite{Harrigan2021}, but is essential for practical application. Similarly, the number of operations run in parallel will decrease the critical path of the circuit and reduce the runtime. With this, both metrics can be abstracted by evaluating the runtime. As quantum computers have progressed to a size that can run practical algorithms such as QAOA and VQE (variational quantum eigensolver), a higher abstraction level using application performance would be the proper method to evaluate the performance of current quantum computers.\\

A very recent benchmark QASMbench \cite{qasmbench} has been presented by Li et al. This benchmark uses the OpenQASM assembly representation and has been used to benchmark several IBM quantum computers. While this benchmark is valuable to benchmark quantum chips, it is too low level to benchmark the quantum computing stack. Understandably, such benchmarks are of use to the development of quantum chips. However, the benchmarking of the full stack will be required with long-term benchmarking in mind. Especially with the current popularity of hybrid quantum algorithms, the quantum chip as well as its classical counterpart and the interaction between them must all be included in performance benchmarking.

Another proposed benchmark is the BB84 protocol, used in quantum communication \cite{BB84_Zukov}. The BB84 protocol uses only two qubits and can therefore be applied to the majority of currently available quantum hardware. It can, however, be argued that the algorithm in itself does not show a large variety in gates and that hardware can be optimized to these operations. It can therefore be considered a useful benchmark, if combined with other quantum algorithms.

A paper by Michielsen et al. discusses various quantum algorithms as candidates for benchmarking \cite{Michiesen_benchmarking}. This set of benchmark problems in its implementation has shown informative results of IBM experience systems, but is again aimed at low level benchmarking rather than the full stack. The algorithms are purely quantum, whereas hybrid algorithms reflect current usage of NISQ quantum computers better, and show more insight to the quantum stack rather than the quantum computing chip in itself. 

Earlier this year, Atos has announced the Q-score benchmark \cite{atos_2020} that evaluates to which problem size a quantum computer can run an optimization algorithm for the MaxCut problem, to which the details are provided in Section \ref{sec:MaxCut}. This QAOA-centric benchmark is a good first approach to establish meaningful quantum benchmarks inspired directly from applications. Atos' Q-score benchmark targets the following metrics:
\begin{itemize}
    \item \textbf{Success-rate, accuracy}: How well is the algorithm executed
    \item \textbf{Performance}: How fast is the algorithm executed
    \item \textbf{Scaling}: How well does the algorithm perform on different problem sizes
\end{itemize}
We believe that further metrics of practical relevance should be included to create an even more insightful benchmark. First, we suggest a more fine-grained splitting of the total runtime into the different stages: (1) compilation, (2) classical computation, (3) communication between classical and quantum and (4) quantum computation. Each of these four stages targets a different aspect of the quantum computing stack and will give both quantum hardware providers and algorithm developers insight into the computational bottlenecks. Furthermore, the importance of measuring runtime should be stressed. Runtime allows for a future proof way of benchmarking, as mature quantum computers will likely be less concerned on the number of qubits, or the quality of qubits when we grow towards perfect qubits. Runtime will become the dominant metric, as when quantum computers become commercially viable, this metric will determine development time and computation cost.

Another concern of the Q-score benchmark is its lack of universality. While it claims to run on every current quantum computer, it only uses a single problem: MaxCut. The danger of using a single problem is that quantum computers could be tailored to this specific problem but perform poorly with others. An extreme case is the D-Wave quantum computer, which is specialized to quantum annealing by design. This is by no means a criticism towards D-Wave, but illustrates how a single-problem benchmark could give an incomplete picture of a quantum computer's performance in general. QAOA is a versatile algorithm that can be applied on a large variation of problems with various practical applications. To avoid single problem tailoring, we suggest multiple different problems to benchmark the quantum hardware. %Further details on the requirements of quantum benchmarks are discussed in Chapter \ref{sec:implementation}.\\

An extension to the volumetric benchmark was proposed by the Quantum Economic Development Consortium (QED-C) in October 2021 \cite{QEDC_benchmark}. This open source benchmark probes a quantum backend with small applications for which the problem sizes are varied, mapping the fidelity of the results as a function of circuit width and depth, hence making it a volumetric benchmark. Other than this primary metric, the benchmark also provides insight for runtime and ratio between programmed and transpiled circuit depth. They make a remark that runtime metrics are currently rudimentary, as quantum providers can have different definitions of quantum execution time. Currently, only the quantum runtime is measured in the QED-C benchmark, and further runtime measurements are mentioned as future work.\\

%% file: Selecting_QA.tex
\section{Near term quantum algorithms}
\label{sec:Quantum_Alg}

In this chapter, a sort review of various hybrid quantum optimization algorithms are given. Hybrid quantum algorithms were chosen as these are promising to be effective for relatively few qubits \cite{Guerreschi_2019}. For this reason, the quantum approximate optimization algorithm (QAOA) and the variational quantum eigensolver (VQE) are briefly explained. The chosen algorithm will be explained in further detail. Finally, emerging trends and concerns on implementing the algorithm (parameter tuning, etc.) will be discussed.

\subsection{Near term hybrid quantum optimization algorithms}
\noindent Quantum algorithms have the potential to outperform classical algorithms, also called quantum supremacy. Many quantum algorithms have been developed and algorithms such as Shor's algorithms are proven to perform better than their classical counterpart. Currently the development of quantum hardware is not yet at the scale to support such algorithms. In order to achieve practical quantum algorithms, with quantum supremacy in mind, algorithms have emerged to fit the current NISQ technology. With this technology, algorithms for 50-100 qubits could be supported.  In this section, a selection of similar near term algorithms are discussed which are candidates for near term implementation. 

\subsection*{Quantum approximate optimization algorithm (QAOA)}
\noindent The first algorithm discussed in this section is QAOA \cite{farhi2014quantum}, introduced by Farhi et al.\ in 2014. This algorithm can be used to approximate a set of NP-hard optimization problems with relatively few qubits. NP-hard optimization problems are a large bottleneck for current classical computing systems, as these cannot be efficiently calculated using classical computers. The type of optimization problems that QAOA can be applied to consist of a set of parameters with a set of constraints where the parameters need a specific configuration for finding the optimal solution. QAOA is a layered quantum algorithm that theoretically increases in accuracy with the increase of the number of iterations (layers) denoted by parameter $p$. This has many practical applications, ranging from multiprocessor scheduling to warehouse-order picking and flight-timetable planning. The specifics of these applications will be further discussed in Section \ref{sec:QTRL}.

\subsection*{Variational quantum eigensolver (VQE)}
\noindent The variational quantum eigensolver (VQE) is a quantum algorithm that finds the eigenvalues of a Hamiltonian $H$, representing the system to be solved \cite{VQE}. It is used to minimize the objective function $\bra{\psi(\theta)}H\ket{\psi(\theta)}$. The minimum eigenvalue sought is represented by the ground state of the system. The state $\ket{\psi(\theta)}$ can be varied by changing $\theta$ to converge to the minimum, by means of a classical optimizer. This is a minimum eigenvalue problem and is a constraint satisfaction problem (CSP), which is generally NP-hard \cite{Osborne_2012}. It is also one of the algorithms, like QAOA, that is believed to be one of the first algorithms to run effectively on a near-term quantum computer.

\iffalse
\subsection*{Quantum adiabatic algorithm (QAA)}
\noindent The quantum adiabatic algorithm (QAA) is able to find the optimal solution to a satifyability problem when given enough time by finding the ground state of the slowly varying Hamiltionian $H(t)$ \cite{QAA}. QAA uses the time-dependent Hamiltonian $H(t) = (1-t/T)B+(t/T)C$. The algorithm starts in the highest energy state of $B$, and seeks the highest energy state of $C$. This makes use of the Perron-Frobenius theorem that the difference in energies between the top state and the one below is greater than 0 for $t<T$ \cite{farhi2014quantum}. This means that if $T$ is sufficiently large, the optimal solution can be found by \textit{slowly} varying $t$ for $0 \leq t \leq T$ \cite{QAA}. A downside of QAA is that success probability is not a monotonic function of $T$. This means that the success probability can drop, whereas for QAOA the success probability always increases with $p$ \cite{farhi2014quantum, crosson2014different}. It is also shown that QAA can be trapped in a local minimum, which is not the case for QAOA \cite{farhi2014quantum}. It is also worth mentioning that QAA is not a gate-based algorithm.\\ 
\fi
%\subsection{Practical concerns and emerging trends for QAOA} %\label{sec:practical}
\subsection{Practical concerns for QAOA}
\noindent In order to apply QAOA algorithms to practical applications, some hard challenges must be achieved first. One of the major challenges is that even while QAOA uses much less resources than, e.g., Grover's algorithm, it still requires hundreds of qubits in order to achieve better performance than classical algorithms. Specifically for QAOA, the depth of a circuit can be a huge bottleneck if it does not make use of an optimized implementation \cite{crooks2018performance}. Furthermore, the limited precision of QAOA will require a large parameter $p$ to achieve optimal results. This parameter is expected to grow almost linearly with the problem size, but finding the optimal value for $p$ is one of the big challenges for QAOA to become practical \cite{Shaydulin_2019}. Another challenge is the tuning of parameters $\bm{\beta}$ and $\bm{\gamma}$. The original paper of QAOA \cite{farhi2014quantum}, proposes an optimal configuration for the $\bm{\beta}$ and $\bm{\gamma}$ parameters, but is considered computationally expensive \cite{crooks2018performance, delagrandrive2019knapsack}. Instead, optimal parameters can be found for specific problem instances \cite{crooks2018performance}. For specific problems, a value for $p$ can be chosen in such a way that the approximation is better than the classical counter part. For example, Crooks shows that for $p \geq 8$, the QAOA for MaxCut could outperform the Goemans-Williamson algorithm \cite{crooks2018performance}, which is currently considered the most efficient classical algorithm for MaxCut. This way, a lower bound can be set on the $p$ parameter.\\

Another challenge is that while QAOA might perform good in theory and on a simulator, on real hardware the performance might be considerably worse \cite{willsch2020benchmarking} or could make an implementation unfeasible \cite{sarkar2020quaser}. The relatively large amount of noise in current quantum hardware results in better performance for algorithms with lower depth \cite{Marshall_2020} and will likely be a strong factor in implementing QAOA in practice.

\subsection{Emerging trends}
An emerging trend is to optimize the parameters for QAOA using machine learning \cite{alam2020accelerating, Khairy2019ReinforcementLF, Wauters_2020}. %Results of Khairy et al. show that training for an 8-qubit system, the training time for the QAOA meta-parameters $\bm{\beta}$ and $\bm{\gamma}$ can take up to 4.98 hours \cite{Khairy2019ReinforcementLF}. While this optimization improves performance, this can only be applied if the algorithm is used many times for the same $n$-bit system in order to achieve speedup.
Results of Khairy et al. show that training for an 8-qubit system, the optimality ratio (equation \ref{eq:opt_r}) can be improved 1.16 to 8.61 compared to the Nelder-Mead optimizer, depending on the MaxCut graph solved and QAOA depth $p$ (depths 1, 2 and 4 were tested).
Wauters et al. \cite{Wauters_2020} aims to solve  this problem by suggesting that a strategy learned from a small system, can be applied to a larger system to achieve similar speedup. Their results using local optimizations on larger systems show this is achievable. Alam et al. \cite{alam2020accelerating} applies machine learning to a classical optimizer that iteratively runs the QAOA algorithm with new parameters. Their implementations show an average reduction of iterations of 44.9\% and is also a good solution for runtime optimization.\\

In terms of integration, Qiskit and Rigetti actively include QAOA in their repositories. Qiskit currently only supports Ising-type problems such as MaxCut \cite{delagrandrive2019knapsack}, while Rigetti also has support for, e.g., the partitioning problem and makes this publicly available on Github.
Rewriting the problem to a format that can be optimized using QAOA is a challenge in itself as well. In recent papers, Quadratic Unconstrained Binary Optimization (QUBO) \cite{glover2019tutorial} is used to map problems to QAOA. QUBO is a compelling approach as this allows the QAOA algorithm to be unaltered, while only adjusting the classical problem formulation. This requires some additional work, but would otherwise be needed to properly design the QAOA algorithm. QUBO is used by, e.g., Qiskit \cite{qiskit} and XACC \cite{XACC}, allowing a single QAOA function definition, with a standardised problem formulation. In terms of performance, no clear comparison has been published comparing the QUBO implementation and the straightforward QAOA implementation.
\\

Another downside of QAOA could be the limited range of problems that it can solve. While all problems discussed in Section \ref{sec:QTRL} are considered important challenges, the quantum algorithm is still limited to very similar problems. A generalization of QAOA is presented as the Quantum Alternating Operator Ansatz, which covers a wider range of problems \cite{Hadfield_2019}. As discussed in Section \ref{sec:QTRL}, some algorithms are not well suited for QAOA and other quantum algorithms will likely perform better results. The constraint density is also shown to impact the performance of QAOA \cite{Akshay_2020}. This means that performance will depend not only on the size of a graph, but also the number of constraints compared to the size of the graph.\\
%penalty terms

%% file: QTRL.tex
\section{Quantum algorithm technology readiness level}
\label{sec:QTRL}

The QAOA implementations of the above discussed problems will be summarized and compared according to their technology readiness level (TRL). This is done, as not every paper provides an in-depth analysis of the proposed implementation. Providing an in-depth analysis of every implementation would be beyond the scope of this paper. TRL was first introduced by NASA \cite{NASA_TRL} to evaluate the state of development of technologies.
The NASA TRL has 9 levels, from proposing basic principals (1) to flight proven (9). Currently, more adaptions have been made for both European projects and commercial projects \cite{Hder2017FromNT}.

\Figure(topskip=0pt, botskip=0pt, midskip=0pt)[clip, trim=0.5cm 0cm 0.5cm 0cm, width=1.5\columnwidth]{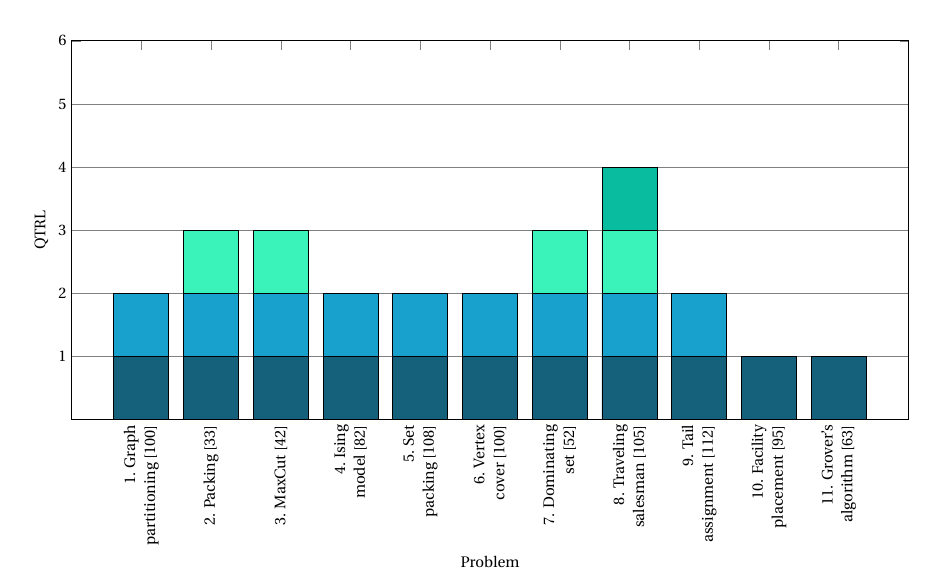}{A quantum technology readiness level (QTRL) evaluation of the presented QAOA implementations.\label{fig:TRL}}

In order to evaluate the maturity of quantum algorithm implementations appropriately, we propose a novel quantum TRL (QTRL). This QTRL is inspired by the TRL as presented by NASA, but adapted to quantum algorithms. The QTRL is in this chapter used to evaluate the presented QAOA implementations, but can be used to evaluate quantum algorithms in general. The QTRL are ordered as follows:\\
%\begin{enumerate}
     \textbf{1) Theoretical proof of concept is presented}: The quantum algorithm to a specific problem is fully worked out and a theoretical proof is given that this implementation approximates or yields the optimal solution to the problem.\\ \textit{In the example of QAOA, this means that both the Cost and Mixer Hamiltonian are presented to the specific problem, as well as proof of the approximation.}\\
     \textbf{2) Demonstrated on quantum simulator}: The quantum algorithm is implemented on a quantum simulator and provides meaningful results. \textit{For QAOA, this means that the simulation should return values that to reasonable extend approximate a valid a solution.}\\
     \textbf{3) Cost and performance analysis}: An in-depth analysis of the implementation is done, showing the required resources and complexity. With resources is explicitly meant: qubits, circuit depth, number of each used gate.\\
     \textbf{4) Simulated application}: The problem is implemented to a practical utility (application) with a classical algorithm that uses the quantum algorithm. The system should work with the combination of a classical computer and a quantum simulator. The full system should provide meaningful/valid results.\\
     \textbf{5) Full stack application}: The full stack is presented with an implementation using quantum hardware and can be applied to problems in practice. The application will likely need optimization or clever implementation to out perform classical implementations.\\
     \textbf{6) Quantum supremacy}: The problem can be solved faster on a classical-quantum hybrid full stack compared to a classical implementation, using the quantum algorithm. \textit{For QAOA, this is currently an aim for the future, as no such implementations have been presented yet. Current efficient algorithms, such as QAOA for MaxCut, still require hundreds of qubits to achieve quantum supremacy. With the current state of the NISQ hardware, systems that can support such large numbers of qubits are not available.}\\
%\end{enumerate}

%\fi

The QTRL of the problems is presented in Figure \ref{fig:TRL}. For each of the presented algorithms, a proof of concept for the QAOA implementation has been presented. Except for the facility placement problem and the QAOA for Grover's algorithms, the algorithms have been tested on a quantum simulator, with published results. On only four of the algorithms, a cost and performance analysis has been published in literature: Packing \cite{setpacking}, MaxCut \cite{farhi2014quantum}, Dominating set \cite{guerrero2020solving} and Traveling salesman \cite{sarkar2020quaser}. These are viable implementations for QPack. We have decided to focus on 3 of the possible canidates as the benchmarks, namely MaxCut problem, Dominating set problem and traveling salesman problem. The packing problem implementation has been left out, due to time constraints and limited availability of implementation details.  Most notably, the Traveling salesman problem has been presented on a full stack, applied to a practical problem \cite{sarkar2020quaser}. It is therefore the only algorithm at time of writing that can be considered QTRL-4. Some notes on the results are:
\begin{itemize}
    \item The MaxCut algorithm has been presented on a full stack quantum computer \cite{Harrigan2021}, but not in the context of a practical application.
    \item Publications on the Ising model have analytic results on complexity, but it is considered not feasible and no resource analysis is done \cite{protein, Akshay_2020}. As such it cannot be considered QTRL-3.
    \item Analysis has been done on the complexity of the set packing implementation, but no in-depth results on resources have been published \cite{ruan2020quantum}.
    \item The traveling salesman has been implemented in full stack and on a practical application, but is not ready to work on available quantum hardware \cite{sarkar2020quaser}.
\end{itemize}

%% file: benchmark_updated.tex
\section{Benchmarking metrics}
\label{metrics}

\noindent To create a standardized benchmark, a set of metrics needs to be agreed upon. In order to properly evaluate the performance of a quantum hardware implementation in practice, the following aspects need to be considered: (1) Runtime, (2) Outcome accuracy and (3) Performance scaling.

\subsection*{Runtime} %%!!! include cite Aritra (discord)
The runtime of an application is an obvious parameter to measure. Classically, apart from the functionality, this is the main parameter measured as it separates implementations on their practical performance. Since current NISQ quantum hardware is only practical in classical-quantum hybrid configurations, both the classical and the quantum hardware need to be benchmarked. While the quantum speedup should be most significant to the performance, for QAOA optimization it was found that the classical optimizer plays an important role to the performance of the QAOA optimization \cite{sarkar2020quaser, bittel2021training}. Determining the performance of the classical hardware is therefore as essential as evaluating the quantum hardware. For this reason the following runtimes need to be observed:
\begin{itemize}
    \item Overall runtime   
    \item runtime classical hardware    
    \item Connection between classical and quantum hardware    
    \item Preprocessing/placing and routing (quantum circuit compilation)
    \item Queuing time
    \item runtime quantum hardware   \\ 
\end{itemize}

The communication between quantum and classical hardware might very well be significant to the overall performance, as QAOA requires many calls to the quantum hardware. Since quantum computing is a Compute as a Service (CaaS) in current implementations, the way the connection is organized as well as the relative slow internet connection can be crucial. In cases that the implementation is organized with a local classical computer sending instructions to the remote quantum computer, most time will be lost by queuing and classical to quantum connection. Queuing time indicates time spend waiting for access to a quantum system. In regular IBM quantum computer calls, the call is put in a queue for access for each quantum circuit call (shots for the same call are executed consecutively). A more mature implementation would be both a remote classical computer and a remote quantum computer, which share a faster local connection. An example of such a quantum stack is the IBM experience runtime. In this case, the classical code is sent in combination with the quantum circuit, as a quantum job. This job in itself is queued, but not each individual call to the quantum computer. Rigetti offers a QaaS as well, but trough an API to log in to their QCS. Rigetti also allows the user to use a remote classical computer to communicate faster to the quantum computer. The service, in contrast to IBM, requires the user to schedule a time slot with access to the quantum computer. This in itself has its pros and cons.\\
By explicitly measuring these detailed timestamps, much more crucial information is learned about the bottleneck of an implementation. As the benchmark does not optimize quantum instructions to suit specific hardware, the performance of the compiler will be reflected as well and will translate to the number of instructions generated and how the quantum circuit is scheduled. Other bottlenecks on the implementation could include: number of calls to quantum hardware from classical hardware and time required per quantum instruction.\\

\subsection*{Outcome accuracy / Best approximation error} 
QAOA can only approximate an optimal solution. While the quality of the approximation depends on parameter $p$ and the classical optimization strategy, it will also depends on the hardware. High quality quantum hardwarew can produce accurate results, but if the optimizer is not accurate the end results will be inaccurate as well, and vice versa. The qubit quality will determine the noise in a system and will determine $p$ \cite{alam2019analysis} which can significantly decrease the runtime performance. Furthermore, gate error and finite coherence time can introduce significant bounds on $p$, limiting not only compensation for noise, but also place significant bounds on scaling. Another effect on the accuracy could be $\ket{0}$ bias \cite{guerrero2020solving}. This means that the states are more likely to be measured as a logical $0$ rather than a logical $1$, skewing the outcome and effecting the accuracy. The approximation can be evaluated in multiple approaches.\\

\noindent
\textbf{Approach 1}:\\ \noindent
The benchmark will be run on the hardware with optimal parameter $p$, determined iteratively. This way the maximum achievable accuracy can be found by comparing it to the solution of an exact algorithm. This has as downside that multiple runs will be required to find the correct parameter, and increasing $p$ will increase the runtime, which might not be properly reflected this way. If the runtime of the implementation is however presented for only the optimal parameter, it will likely better reflect how the optimization would be implemented in practice. This way a decision will be made for the trade-off between speed and accuracy, favoring accuracy. The difficult part is that, in the case that there are no bounds on $p$, $p$ might grow to infinity, approaching exact calculation. Recent research show that increasing $p$ will not always improve accuracy \cite{niu2019optimizing} and will have a practical limit.\\
    
\noindent    
\textbf{Approach 2}:\\ \noindent
The benchmark will be run on hardware for, e.g., $p \in \{1,2,3,4\}$ and the results will be compared. This way the runtime will not be determined by the choice of $p$, but the accuracy will reflect the quality of the quantum hardware. This will require more interpretation, as the accuracy shown will not be the optimal accuracy, but the accuracy that is achieved for low $p$. The difficulty in this approach lies in determining how $p$ scales for larger problem sizes. The size of $p$ is expected to grow with the problem size, but at which rate is not yet determined. Some research suggests that QAOA will only be practical for low depth (e.g. $p$=2) which might eliminate this issue.\\
From both approaches, the latter appears to be the most fair and with fewer issues, but demands an explanation on how to interpret the results.\\

\noindent
\textbf{Approach 3}:\\ \noindent
Alternatively, an approach can be taken where the accuracy is evaluated for a predetermined runtime. This however requires that the runtime will not be constant for each problem size as the runtime is expected to grow. Determining a starting point for the run time as well as how this will progress for the problem size is nontrivial.\\

\noindent
Of the three presented approaches, approach one is preferred.

%\textbf{Success probability}: The probability of success is another measure of the quality of the quantum hardware. Preferably, the best approximation error of the QAOA implementation reaches $0$ for some amount of runs. The probability of success can be determined by the percentage of runs that achieve the optimal solution and will, much like the best approximation error, reflect the quality of the quantum hardware. This method however requires that the implementation can in fact find the optimal solution, rendering the previous method obsolete.

\subsection*{Performance scaling}
With concerns as bounds on $p$ and scaling issues, it is important to give insight on how the benchmark performance scales for larger problem sizes. By determining the runtime for increasing problem size, the scaling performance can be reviewed. Similarly, it can be evaluated on how the accuracy or success probability decreases for larger problem sizes.

While this benchmark is focused on quantum hardware, it could be extended to quantum simulators, evaluating runtime and memory consumption. This could give valuable insight to developers to find the optimal simulator to work with.\\

Benchmarking modern high performance computing architectures has a large focus on energy consumption. This is expected become an important metric for quantum computing in the future as well. Energy consumption is, however, complex to benchmark as most of the measurements need to be done on location, and the division over classical and quantum computing adds another layer of complexity. In current quantum hardware there is a large variety of hardware implementation, making it even more difficult to standardize the measurement of energy consumption. For obvious reasons, this metric is not included in the benchmark, but quantum hardware developers are encouraged to accompany benchmark result with any estimate of energy consumption.

\section{Measurements}
\label{sec:measurements}
In this section, measurements of the different classical optimizer options and the different QAOA implementations are presented. These measurements were done using Qiskit qasm simulator, to show relative performances of classical parameter optimizers and QAOA applications.

\subsection{Measurements of classical optimizers}\label{sec:c_opt_meas}

\noindent In Appendix \ref{sec:param_opt}, a preliminary study of various classical optimizers was done. In the current section, various promising classical optimization algorithms will be applied for increasing problem size to compare the runtime performance and accuracy. Apart from Nelder-Mead and BFGS, which are both discussed in Section \ref{sec:param_opt}, other optimization algorithms available in open-source libraries are presented as well. The list of tested optimizers is:

\begin{enumerate}%[label*=\arabic*.]
    \item Local optimizers
    \begin{enumerate}%[label*=\arabic*.]
        \item Nelder-Mead (nm)
        \item BFGS (Broyden–Fletcher–Goldfarb–Shanno)
%        \item NEWUOA (New Unconstrained Optimization Algorithm) \cite{powell2006newuoa}
%        \item BOBYQA (Bound Optimization By Quadratic Approximation) \cite{powell2009bobyqa}
        \item COBYLA (Constrained Optimization By Linear Approximation) \cite{powell1994direct}
    \end{enumerate}
    \item Global optimizers
    \begin{enumerate}%[label*=\arabic*.]
%        \item Dual annealing \cite{xiang1997generalized}
        \item SHGO (simplicial homology global optimization) \cite{endres2018simplicial}
%        \item ISRES (Improved Stochastic Ranking Evolutionary Strategy) \cite{runarsson2005search}
%        \item MLSL / MLSL\_LDS (Multi-Level Single-Linkage, low-discrepancy sequence) \cite{MLSL}
%        \item DIRECT (DIviding RECTangles) \cite{jones1993lipschitzian}
%        \item StoGO (Stochastic Global Optimization) \cite{zertchaninov1998c++}
    \end{enumerate}
\end{enumerate}

These algorithms were implemented in Python using either SciPy or the NL-Opt library \cite{NLOpt}. %Some of these algorithms are not included in the following results, as either no convergence was found (due to various reasons) or if the optimization had such a long runtime that they cannot be considered a contender in comparison to other algorithms. The algorithms included in this list are: \\

%\begin{itemize}
%    \item Dual annealing
%    \item NEWUOA
%    \item BOBYQA
%    \item ISRES
%    \item MLSL / MLSL\_LDS
%    \item DIRECT
%    \item StoGO
%\end{itemize}

%Perhaps with different implementations, these algorithms could be used for QAOA optimization. In the case of dual annealing, it was clear that convergence could be found, but takes significantly longer than other tested optimization algorithms. The MLSL implementation suffers from searching for every local optima (which can be very significant with noisy QAOA results). A variant of MLSL might work for QAOA, but with the current implementation no convergence could be found in reasonable time.

%\begin{figure}[h]
%    \centering
%    \includegraphics[width=0.9\linewidth]{figures/timing_data.pdf}
%    \caption{Data of runtime measurements fitted to exponential functions. Any number in the legend indicates the convergence tolerance, which is default if not indicated.}
%    \label{fig:timing_opt}
%\end{figure}

\Figure[t!](topskip=0pt, botskip=0pt, midskip=0pt)[width=0.9\linewidth]{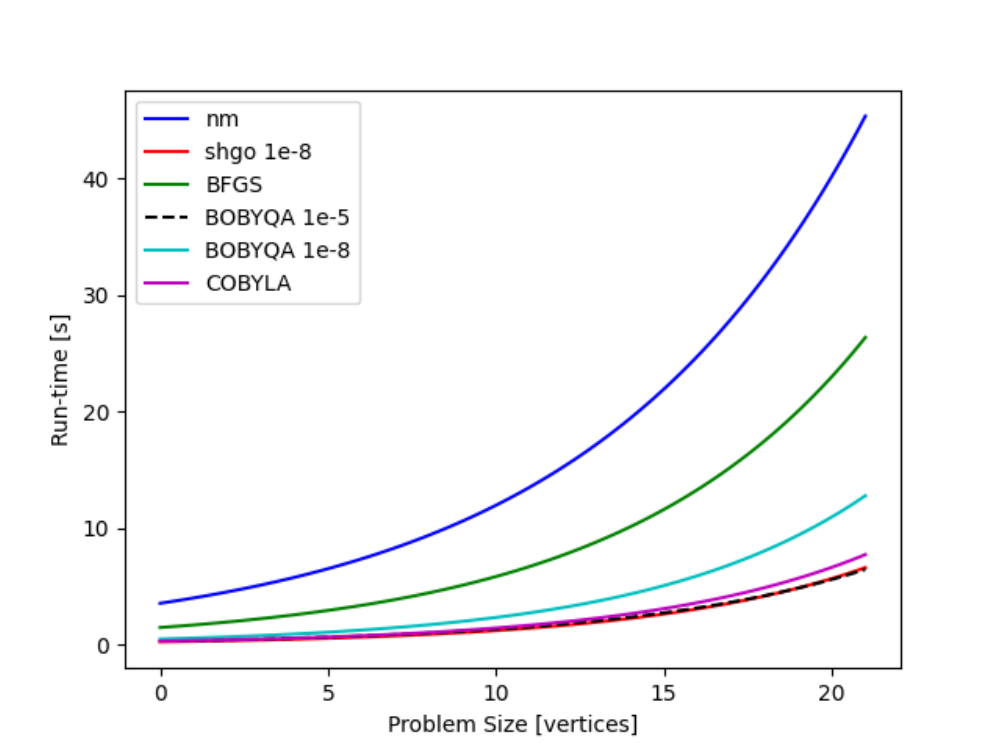}
{Data of runtime measurements fitted to exponential functions. Any number in the legend indicates the convergence tolerance, which is default if not indicated.\label{fig:timing_opt}}

The runtime measurements were done on MaxCut problems with 3 to 23 nodes. The QAOA algorithm is implemented using Qiskit for the  IBM Qasm simulator \cite{quasm} The MaxCut problems are randomly configured, but have the same number of edges as nodes to maintain the problem complexity. The measured runtimes are shown in Figure \ref{fig:timing_opt}. The data is fitted to an exponential function, as the obtained data shows a clear exponential growth. In this figure, it can be seen that the Nelder-Mead algorithms which was expected to perform best, is in fact the slowest from the selected algorithms in practice. SHGO, COBYLA and BOBYQA perform much better in terms of runtime, with SHGO being the fastest. \\

%\begin{figure}
%    \centering
%    \includegraphics[width=0.9\linewidth]{figures/relative_acc.pdf}
%    \caption{Optimizer accuracy compared to Dual-Annealing for the same MaxCut problem instance. Here a lower relative accuracy shows a more accurate algorithm. A negative value shows that the tested algorithm outperforms the Dual-Annealing algorithm. R COBYLA here indicates the COBYLA algorithm with multiple randomized starting points.}
%    \label{fig:acc_opt}
%\end{figure}

\Figure[t!](topskip=0pt, botskip=0pt, midskip=0pt)[width=0.9\linewidth]{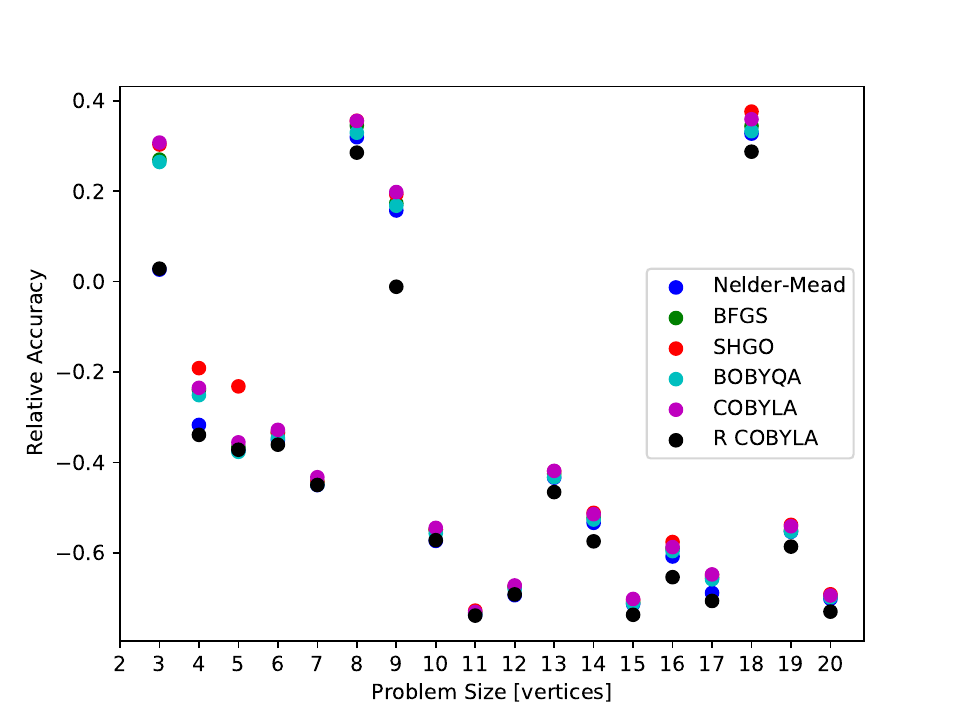}
{Optimizer accuracy compared to Dual-Annealing for the same MaxCut problem instance. Here a lower relative accuracy shows a more accurate algorithm. A negative value shows that the tested algorithm outperforms the Dual-Annealing algorithm. R COBYLA here indicates the COBYLA algorithm with multiple randomized starting points.\label{fig:acc_opt}}

For the accuracy measurements, the different algorithms solve the same MaxCut problem for different sizes and the best expectation values are compared. Preferably, a fine grid search (brute force) is implemented as reference, but this proved to be too time consuming to collect data, even for small problems. In Figure \ref{fig:acc_opt}, the found expected values are compared to the value found by the Dual-Annealing algorithm. Dual-Annealing was chosen, since it was a slower global optimizer, but allowed for reasonable computing times and expecting that its accuracy would outperform the faster algorithms. The relative accuracy (RA) is calculated as 
\begin{equation*}
    \textit{RA} = \frac{\text{\textit{expectation value}} - \text{\textit{measured expectation value}} }{\text{\textit{DA expectation value}}}
\end{equation*}

The results however show that Dual-Annealing does not always find better solutions and is often outperformed by other contenders. The results show clearly that COBYLA using randomized starting points outperforms other algorithms, indicating that a hybrid approach finds the best solutions. This, however, significantly increases the search time. For the single algorithm approaches, Nelder-Mead finds the best expectation values, presenting a trade-off between accuracy and time. It should be noted that for larger problem sizes, the expectation values are much closer and it is expected that for scaling purposes it is better to choose a faster algorithm such as COBYLA, BOBYQA of SHGO. Some parameters such as the number of the QAOA iterations or the convergence tolerance are not optimized as this is expected to give minimal change to the results with respect to the comparison.\\

Considering the found results in terms of accuracy and runtime, a hybrid approach could be applied. Heuristics such as random starting points combined with a local optimizer are often applied, to counteract the local optimizer of getting trapped in local optima. Another implementation is the use of a (fast) global optimizer with a more accurate local optimizer. In the case of optimizing QAOA parameters, using a fast algorithm such as SHGO combined with, e.g., Nelder-Mead or BFGS would be an interesting approach.  This approach however, did not provide enough evidence to indicate better performance compared to random starting points. The increase of starting points in itself proved to increase the accuracy, but the method did not affect the accuracy significantly.

%%%%%%%%%%%%%%%%%%%%%%%%%%%%%%%%%%%%%%%%%%%%%%%%%%%%%%%%%%%%%%%%%%%%%%%%%%%%%%%%%%%%%%%%%%%%%%%%%%%%%%%%%%%%%%%%%%%%%%%%%%%%%%%%%%%%%%%%%%%%%%%%%%%%%%%%%%%%%%%%%%%%%%%%%%%%%%%%%%%%%%

\subsection{Measurements of applications} \label{sec:app_measurements}
\noindent A common downside of computer benchmarks is that hardware vendors are tempted to tailor their systems to excel in the particular benchmarks. We have therefore included multiple applications to the QPack benchmark. Next to the MaxCut problem (MCP), the dominating set problem (DSP) and the traveling salesman problem (TSP) have been included. In this section, the details on the QAOA implementation will be discussed, as well as the effect on the measured performance.
The code implementations of all problems can be found on the QPack github repository, listed in section \ref{sec:code_repos}.\\

\noindent
\textbf{MaxCut}\\
The MaxCut application has been examined in detail in Section \ref{sec:MaxCut}. The QAOA implementation requires the following gates for $n$ vertices and $m$ edges(or clauses) and $p$ QAOA iterations:
\begin{table}[h]
\centering
\begin{tabular}{|c|c|}
\hline
    Gate type  & \# gates\\
    \hline
    Hadamard & $n$ \\
    CNOT & $2m \cdot p$ \\
    RZ & $m \cdot p$ \\
    RX & $n \cdot p$ \\
    \hline
\end{tabular}\\
\end{table}

From the applications explored, MaxCut QAOA requires the least gates and is therefore expected to run best on NISQ devices. MaxCut results in the smallest circuit depth and does not require additional ancilla qubits.\\ 
%The implementation of the MaxCut circuit and the cost function evaluation are presented in Appendix \ref{app:mcp-qiskit}.\\

\begin{figure}[h]
    \centering
    \includegraphics[clip, trim=5cm 8.5cm 5cm 3cm, width=0.5\linewidth]{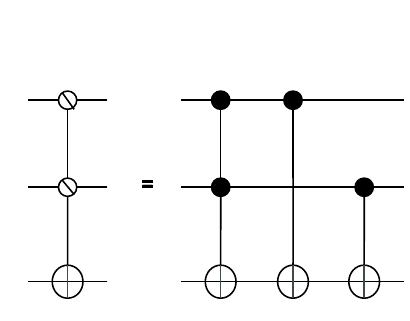}
    \caption{Implementation of a 2-input logical OR using CNOT gates.}
    \label{fig:2OR}
\end{figure}

\noindent
\textbf{Dominating set}\\
The QAOA for DSP is implemented according to the algorithm provided by Nicholas Guerrero \cite{guerrero2020solving}. This implementation uses logical OR gates, which can be implemented using CNOT gates and ancilla qubits. The number of required ancillas and CNOT per logical OR depends on the number of input qubits $k$. The number of required ancillas is $k-1$. The implementation by Guerrero uses the logical OR to control a RY gate. This OR-controlled RY gate requires one additional ancilla as target qubit. As the ancillas can be re-used, the ancillas per circuit add up to $k$ ancillas, with $2 \leq k \leq n$. The logical 2-OR uses 3 CNOT gates, as depicted in Figure \ref{fig:2OR}. These 2-OR gates can be combined to a $n$-input logical OR. %For $n$ inputs, this $n$-OR is implemented as: \\
\iffalse
\lstset{
  basicstyle=\footnotesize, frame=tb,
  xleftmargin=.01\textwidth, xrightmargin=.2\textwidth
}
\begin{lstlisting}
qc = QuantumCircuit(2*n)
qc.append(ORGate, [0, 1, n])
for i in range(2, n):
    qc.append(ORGate, [i, n+i-2, n+i-1])
qc.crz(gamma, 2*n-2, 2*n-1)
for i in range(n, 2, -1):
    qc.append(ORGate, [n-i-1, 2*n-2-i, 2*n-1-i])
qc.append(ORGate, [0, 1, n])

\end{lstlisting}
\fi

The circuit can be implemented using the following resources (at most):
\begin{table}[h]
\centering
\begin{tabular}{|c|c|}
    \hline
    Gate type &  \# gates\\
    \hline
    Hadamard & $n$ \\
    X & $2m \cdot p$ \\
    CNOT & $((6n-5) \cdot m) \cdot p$ \\
    RZ & $m \cdot p$ \\
    RX & $n \cdot p$ \\
    \hline
\end{tabular}\\
\end{table}

$n$ qubits are initialised in superposition, using $n$ Hadamard gates. Inverted controlled RZ-gates and OR-controlled RZ-gates are used to implement the $m$ clauses \cite{guerrero2020solving}. The $n$ RX gates are then applied to the $n$ qubits. Excluding the initialization, this is repeated for $p$ iterations.

The DSP QAOA implementation uses similar, but significantly more resources than the MCP implementation. The DSP implementation uses $(6n-7)\cdot m \cdot p$ more CNOT gates and $2n \cdot p$ additional Pauli-X gates. This requires quantum computers to support larger circuit depth, but also more ancilla qubits and better CNOT mapping. The additional CNOT requirement will reflect the qubit connectivity of the quantum hardware implementation, as well as the handling of qubit SWAPS (if required) by the hardware scheduler.\\ 
%The code implementation for the DSP circuit and cost evaluation are given in Appendix \ref{app:dsp-qiskit}, again using Python and qiskit. Some of the larger gates, which are not native to qiskit, are omitted in the code below. These are presented in Appendix \ref{app:a}.\\

\noindent
\textbf{Traveling salesman}\\
The TSP QAOA implementation is a much more demanding quantum algorithm and with the current implementation also uses different resources. The implementation is done according to the github %\textbf{\texttt{\url{https://lucaman99.github.io/new_blog/2020/mar16.html}}}
\cite{tsp_git}, which is one of the few examples of a clear implementation. The author however indicates that the state preparation is sub optimal and is therefore altered to suit the algorithm better. The TSP algorithm, as other QAOA algorithm, consists of a state preparation, a mixer Hamiltonian and a cost Hamiltonian. Whereas the MCP and DSP QAOA need $n$ to $2n$ qubits for $n$ vertices, this implementation uses $n^2$ qubits. Other variations exist, such as presented by Ruan et al. \cite{ruan_tsp}, using $m$ qubits, which is at most $n(n-1)/2$ qubits for a fully connected graph. This implementation will be considered a future improvement. This however means that the 
current implementation requires significantly more qubits compared to MCP and DSP. \\

%\begin{figure}[h]
%    \centering
%    \includegraphics[width=0.9\linewidth]{figures/singleplot_app.pdf}
%    \caption{Runtime comparison for MCP, DSP and TSP QAOA applications, implemented in Qiskit, plotted in a single figure (100 QAOA runs were done for comparison).}
%    \label{fig:app_runtime}
%\end{figure}

\Figure[h](topskip=0pt, botskip=0pt, midskip=0pt)[width=0.9\linewidth]{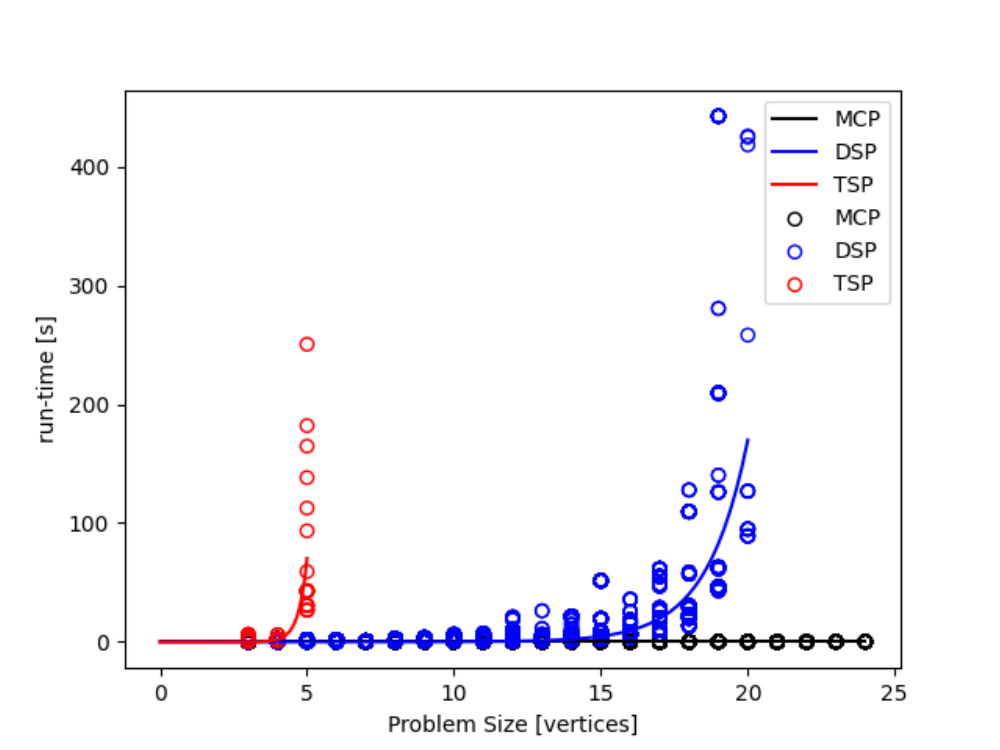}
{Runtime comparison for MCP, DSP and TSP QAOA applications, implemented in Qiskit, plotted in a single figure (100 QAOA runs were done for comparison). \label{fig:app_runtime}}

\begin{figure}[h]
    \centering
    \includegraphics[width=0.9\linewidth]{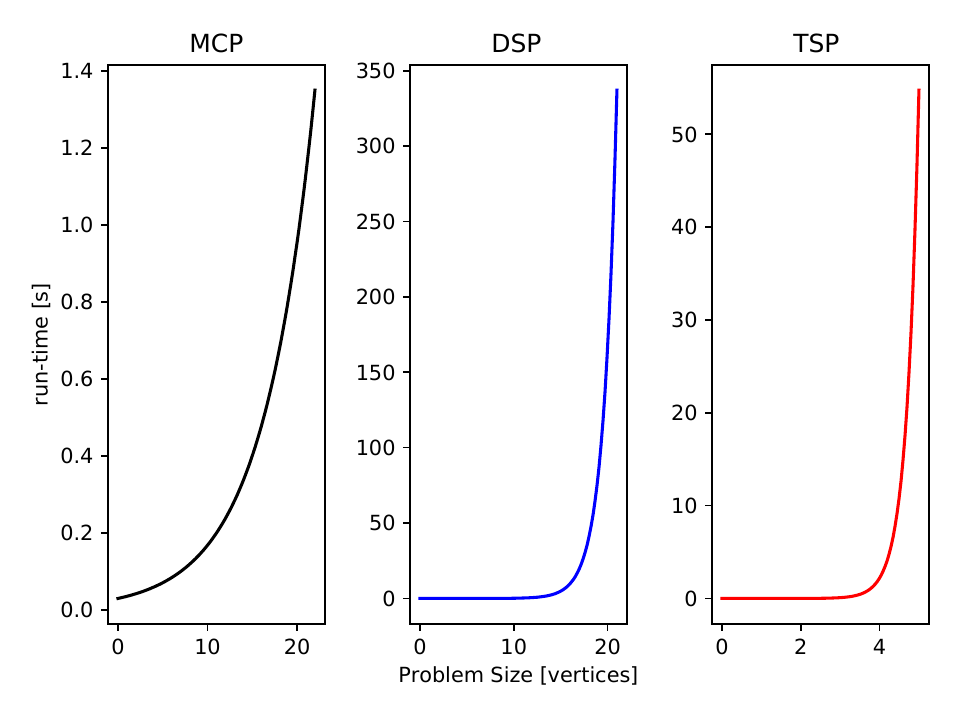}
    \caption{Runtime comparison for MCP, DSP and TSP QAOA applications, implemented in Qiskit plotted in separate figures for detail.}
    \label{fig:app_runtime_2}
\end{figure}

For the state preparation $n$ binary strings of length $n$ are required with a Hamming weight of 2. The superposition of all possible strings with a set Hamming weight is called a Dicke state \cite{Dicke}. In the TSP implementation a Dicke state preparation is applied on each row of $n$ qubits for Hamming weight 2, using the implementation by Bärtschi and Eidenbenz \cite{Dicke_state_prep}. The cost and mixer Hamiltonian are left unchanged. Considering the significant cost of both the Dicke state preparation and the QAOA cycles, it worth mentioning the costs separately. The gate counts are provided in the following table: \\

\iffalse
\begin{multicols}{2}
\centering
\begin{tabular}{|c|c|}
\hline
 \multicolumn{2}{|c|}{Dicke State Preparation Cost} \\
    \hline
    Gate type &  \# gates\\
    \hline
    Hadamard & $8n(n-2)$ \\
    $T/T^\dagger$ & $49n(n-2)$ \\
    CNOT & $(31n-28)n$ \\
    RY & $(6n - 10)n $ \\
    X & $2n$ \\
    \hline
\end{tabular}

\centering
\begin{tabular}{|c|c|}
\hline
 \multicolumn{2}{|c|}{TSP QAOA cycles Cost} \\
    \hline
    Gate type &  \# gates\\
    \hline
    RZ & $n^2 \cdot p$ \\
    RZZ & $\frac{n^2-n}{2} \cdot p$ \\
    RXX & $2n \cdot p$ \\
    RYY & $2n \cdot p$ \\
    \hline
\end{tabular}
\end{multicols}
\fi

\noindent \begin{tabular}{|c|c|c|c|}
\hline
 \multicolumn{2}{|c|}{Dicke State Preparation Cost}&   \multicolumn{2}{|c|}{TSP QAOA cycles Cost}\\
    \hline
    Gate type &  \# gates &  Gate type &  \# gates\\
    \hline
    Hadamard & $8n(n-2)$ & RZ & $n^2 \cdot p$ \\
    $T/T^\dagger$ & $49n(n-2)$ &  RZZ & $\frac{n^2-n}{2} \cdot p$ \\
    CNOT & $(31n-28)n$ &     RXX & $2n \cdot p$ \\
    RY & $(6n - 10)n $ &     RYY & $2n \cdot p$  \\
    X & $2n$ & & \\
    \hline
\end{tabular}\\\\

This shows that while not only the circuit cost is significantly higher, many different gates are used. This will avoid fine tuning of quantum computers on solely Hademard, CNOT, RX and RZ Gates. To compare the impact of the different applications, a runtime comparison was done. The applications are implemented using the SHGO classical optimizer. The results in Figures \ref{fig:app_runtime} and \ref{fig:app_runtime_2} show that the runtime corresponds to the increase of circuit depth and required qubits as expected. Note that Figures \ref{fig:app_runtime} and \ref{fig:app_runtime_2} do not show large problem sizes for DSP and TSP as the larger problem sizes result in simulator memory issues. Despite the smaller data range, the difference in runtimes can be clearly distinguished. 
%The code implementation of the TSP circuit as well as the cost evaluation, are given in Appendix \ref{app:tsp-qiskit}. The implementation is again in Python and qiskit and omitted gates are given in Appendix \ref{app:a}

%\cite{Zhou_2020}: NM slightly outperforms BFGS, even with initial point generation
%EDIT: NM is NOT global, but generally performs better with few local minima

\section{Implementation}
\label{sec:implementation}
%figure: https://arxiv.org/pdf/1812.01041.pdf
Our QPack benchmark provides reference implementations for the aforementioned problems, can be run for increasing problem size, until the quantum hardware is unable to find a solution. The QAOA runs themselves will need multiple runs to determine the average runtime and accuracy. The solutions found by the QAOA optimization, need to be compared to a pre-computed answer using a classical exact algorithm. If the hardware supports larger circuit depth, the QAOA algorithm can be adjusted to a maximum $p$. As theoretically the best solution is always larger or equal to solutions achieved with smaller QAOA depth $p$, a maximum value can be found when the found solution does not improve or diminish. Diminishing of the solution indicates that no larger circuit depth is possible. This search to a maximum $p$ can be included as an additional step for the benchmark to find the best accuracy of the QAOA implementation. The maximum circuit depth and width of quantum hardware will translate to maximum resources per available memory for a quantum simulator. \\

%To give insight on $\ket{0}$ bias, a trivial circuit can be run beforehand. The measurements will indicate if there is a bias. The implementation of QAOA will be done using a classical optimizer which is further explored in Section \ref{sec:c_opt_meas}.

As the benchmark needs to be reproducible, the configuration of the problems must be predetermined. Akshay et al. \cite{Akshay_2020} show that the performance of QAOA in terms of configurations is mainly due to the density (e.g. in a graph: the number of edges of a graph relative to the number of vertices). As the benchmark needs to scale to any problem size, the problems configurations cannot be hard-coded. In the proposed benchmark, the density will remain the same, to make the performance predetermined.

%\begin{figure}[h]
%    \centering
%    \includegraphics[width=0.5\linewidth]{figures/reg_graph.png}
%    \caption{Structure of the circular 4-regular graph. Each node is connected to its neighbour, as well as the node %next to the neighbours. The figure shows an example of a 7-node graph for such a structure.}
%    \label{fig:reg_graph}
%\end{figure}

\Figure[ht](topskip=0pt, botskip=0pt, midskip=0pt)[width=0.5\linewidth]{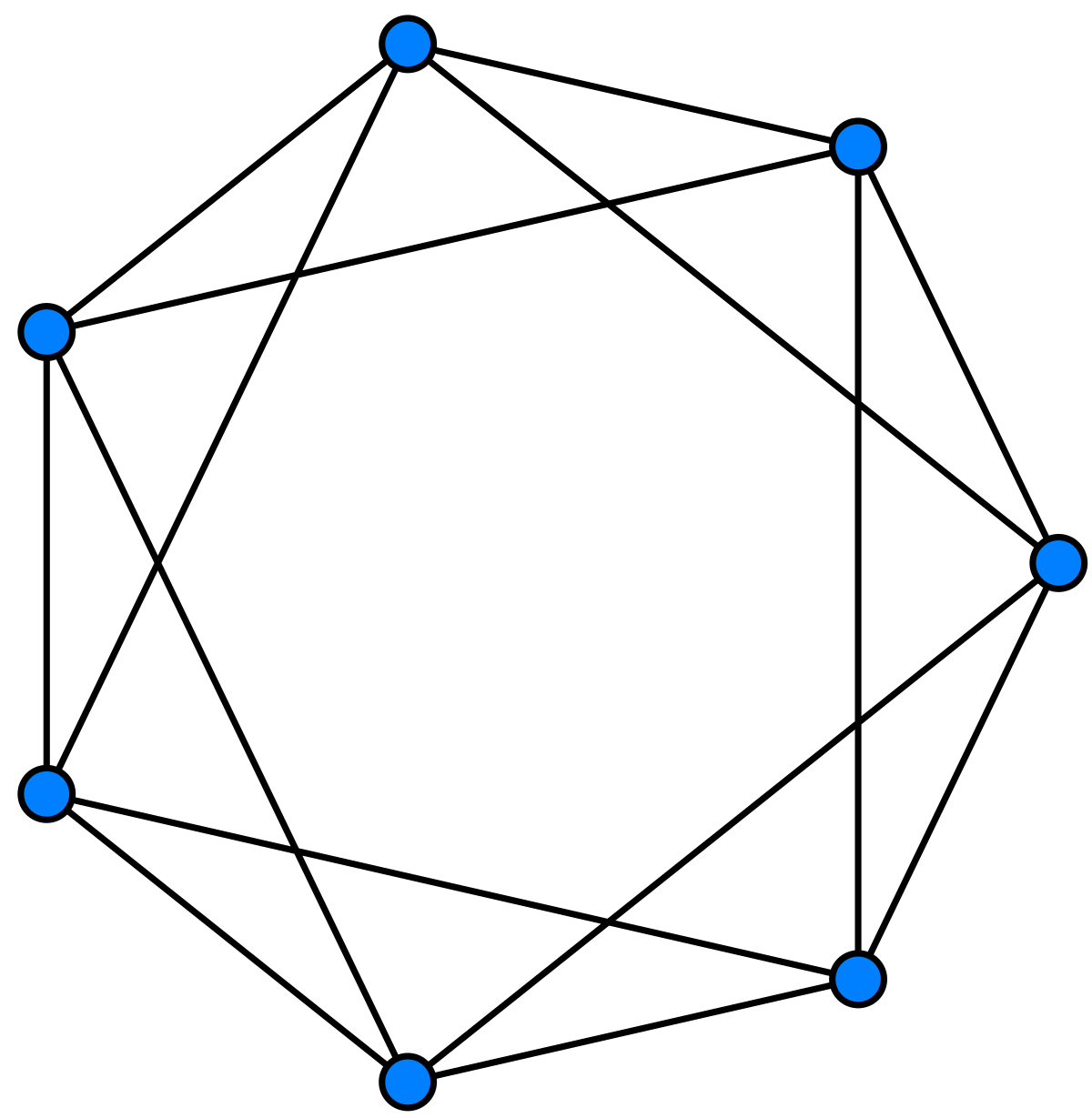}
{Structure of the circular 4-regular graph. Each node is connected to its neighbour, as well as the node next to the neighbours. The figure shows an example of a 7-node graph for such a structure.\label{fig:reg_graph}}

In order to create a scalable problem set with continuous edge density, a 4-regular graph will be used. Specifically, the 4-regular graph will have a structure similar to Figure \ref{fig:reg_graph}. This however is not necessarily representative of practical applications, but will ensure reproducability and scalability without requiring score adjustments to random generated graphs.

The number of "shots" for a single QAOA iteration will range between 1000 and 10000, as these give the best results \cite{Zhou_2020}. The number of iterations to achieve an estimation of the accuracy must be further analyzed. Expected is that the estimation will saturate towards a maximum reliability. A lower bound must be determined for this reliability. Determining the parameter $p$ is also not trivial. There are arguments that the accuracy of QAOA will not grow for $p>2$ \cite{Marshall_2020}, but others argue that e.g. a minimum of $p \geq 8$ is required for the MaxCut application to compete with classical approaches \cite{crooks2018performance}. Determining $p$, possibly dependent on the application will require further numerical analysis.

\begin{figure}[ht]
    \centering
    \includegraphics[trim={1.4cm 5.5cm 5.5cm 3cm},clip, width =\linewidth]{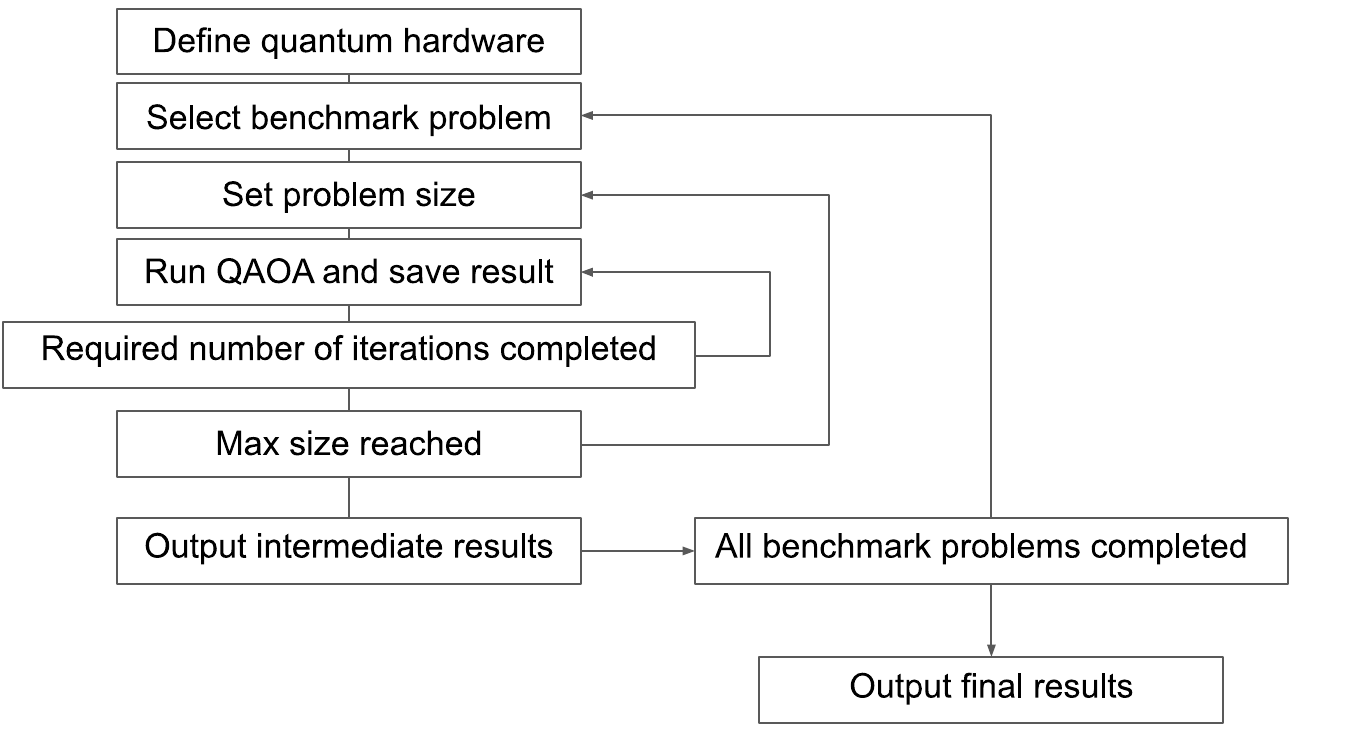}
    %\vspace{-2cm}
    %\hspace{-3cm}
    \caption{Flowchart for the proposed QPack benchmark.}
    \label{fig:flowchart}
\end{figure}

The outline for QPack is given schematically in Figure \ref{fig:flowchart}. Here, the QAOA applications are run for increasing problem sizes. The benchmark will stop if for all applications the maximum problem size has been reached. The maximum is determined when not enough qubits are available to evaluate the problem. Some design choices can be made to make sure that the benchmark is as complete as possible. First of all, the generation of problems can be done every iteration of the QAOA optimization. It would be fairest to create a new configuration of the problem, as certain configurations might be easier to optimize and might give an unfair advantage. This however will require longer computation times for the benchmark itself. Including a random configuration for each run however, might cause benchmark result to become incomparable, and a significant amount of time will be required to validate each generated graph. The number of QAOA runs is also debatable. More runs will give a better average, but a limit must be set to limit the runtime. Another interesting addition would be adding both weighted and unweighted problem instances for all benchmarks. This can show if the hardware can adjust to weighted edges with e.g. using parameter "sensitive" edges \cite{entropica}. This solution would preferably be implemented by the benchmark, but requires too much information about the hardware, which should be abstracted. A quantum instruction compiler would be tasked to implement the qubit SWAPs required for this method. A further concern is that the benchmark should allow for parallelism. With the current size of quantum hardware, it is unlikely that this will be fully exploited in practice. However, as quantum hardware will scale up, there should be support to apply the parallelism. For one of the viable optimizations algorithms, the Nelder-Mead classical optimization algorithm, a parallel variant has been developed \cite{ozaki2019accelerating, dennis1988parallel, lee2007parallel}. This will allow the use of more quantum resources to compute smaller problems. The implementation of such a parallel implementation should be considered as quantum hardware can provide the required resources.

Finally, the implementation of the benchmark will require universality, in the sense that it should run on every quantum computer. The LibKet library \cite{LibKet} is suited for this task. It will, however, take more work to fully adjust the benchmark to this library. Another option would be the XACC library \cite{XACC}, which is conveniently targeted towards hybrid computing. This library provides both a Python and a C++ implementation. XACC supports IBM, Rigetti, IonQ and D-Wave QPUs (and several quantum simulators), whereas LibKet has larger support but is not fully developed for hybrid quantum computing yet. Using this implementation, several quantum simulators have been benchmarked which will be presented in Chapter \ref{sec:results}.

To give the reader an idea of using the qiskit implemented benchmark. a code fragment of the benchmark is given below. For this code fragment all account details should be filled by the user, as well as the backend they would like to use. The function \textit{Benchmark} allows the entries \textit{'mcp'}, \textit{'dsp'} and \textit{'tsp'} for initializing the different benchmarks. The function \textit{set\_lim(int)} sets the largest problem instance size the benchmark will run. The function \textit{set\_p(int)} will set the number of QAOA iterations $p$, which defaults to one. Finally, the function \textit{run(void)} will run the benchmark for increasing problem instance sizes up to a (user defined) limit. The results are then presented in JSON format.

%\begin{lstlisting}
%IBMQ.load_account()
%provider = IBMQ.get_provider(hub='ibm-q', group='open', project='main')
%backend_tag = 'ibmq_qasm_simulator'
%backend = provider.get_backend(backend_tag)

%#MCP benchmark
%mcp = Benchmark('mcp')
%mcp.set_backend(backend, backend_tag)
%mcp.set_lim(10)
%mcp.set_iter(1000)
%mcp.update_p(2)
%mcp.run()
%\end{lstlisting}

%% file: results.tex
%\section{Results}

%This section covers the results of QPack on various quantum hardware and simulators. An qiskit implementation is used to show results on IBM hardware and simulators. Furthermore, a XACC implementation is presented, showing measurements on various simulators. This implementation demonstrates the universality of QPack.

\section{Benchmark results}
\label{sec:results}

\begin{figure}[h]
    \centering
    \includegraphics[width=\linewidth]{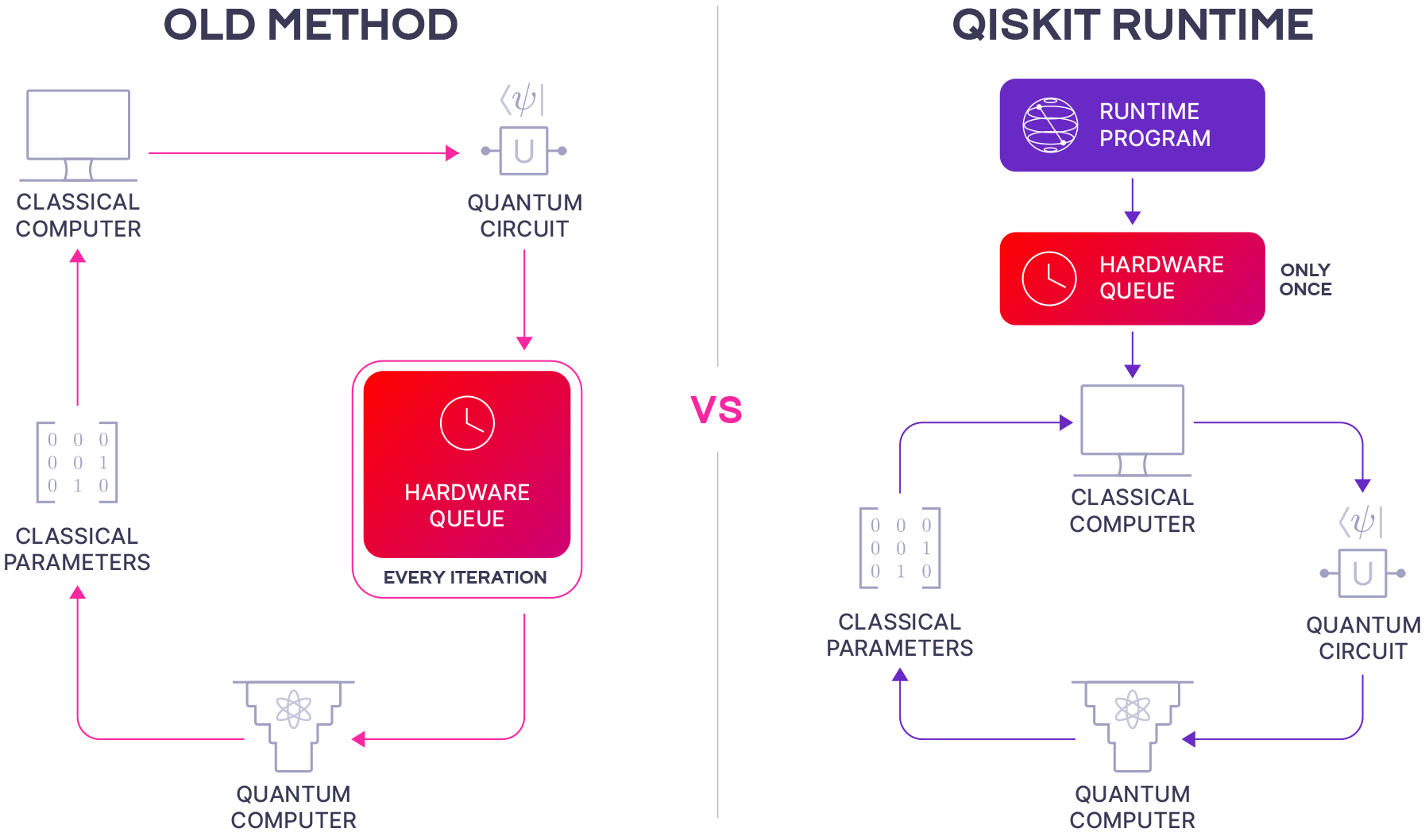}
    \caption{Schematic of the IBM runtime environment, an example of using remote classical computation for quantum computing \cite{strangeworks_runtime}.}
    \label{fig:runtime_env}
\end{figure}

In this section the results are presented for each benchmark discussed in this paper. The results are shown for the remote IBM QASM simulator (32 simulated qubits), IBM Montreal (27 physical qubits, Figure \ref{fig:qubit_montreal_nairobi}) and IBM Nairobi (7 physical qubits, Figure \ref{fig:qubit_montreal_nairobi}). For the IBM Nairobi, the IBM runtime environment is available, which supports the use of a local classical computer, in order to speed up communication for hybrid computing (Figure \ref{fig:runtime_env}). Currently, IBM runtime environment is available for all their hardware, but at the time of testing it was only available for the IBM Nairobi. %Interestingly, the benchmark proposed by the QED-C also mentions this feature, but did not implement any support \cite{QEDC_benchmark}. 
Both the score and runtime results are presented in their respective sections. The results have been obtained using $1000$ shots per iteration, with the number of function evaluations of the classical optimizer limited to $1000$. The results obtained are scaled up from problem size 5 up to a larger size that still allows a reasonable runtime. In the following, we first discuss the runtime, the accuracy and then the scalability of the IBM hardware, followed by the XACC results simulated on multiple hardware platforms.

%\begin{figure}[h]
%\begin{subfigure}[b]{0.20\textwidth}
%    \includegraphics[height=2cm]{figures/qubit_montreal.png}
%    \caption{Qubit layout of the IBM Montreal, as presented on IBM experience.}
%    \label{fig:qubit_montreal}
%  \end{subfigure}
%  \begin{subfigure}[b]{0.20\textwidth}
%    \includegraphics[height=2cm]{figures/qubit_nairobi.png}
%    \caption{Qubit layout of the IBM Nairobi, as presented on IBM experience.}
%    \label{fig:qubit_nairobi}
%  \end{subfigure}
%  \end{figure}

\Figure[h](topskip=0pt, botskip=0pt, midskip=0pt)[width=0.9\linewidth]{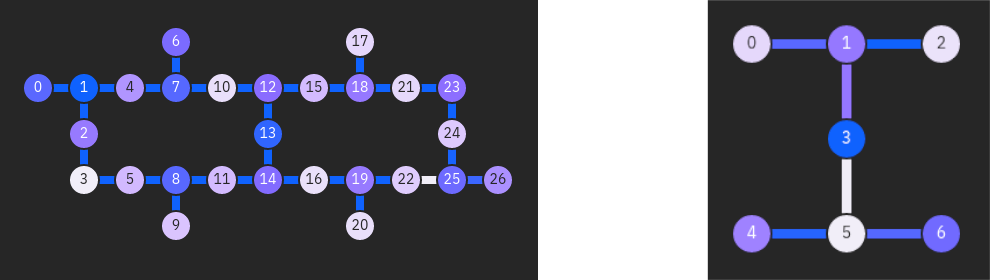}
{Left: Qubit layout of the IBM Montreal, Right: Qubit layout of the IBM Nairobi. As presented on IBM experience \label{fig:qubit_montreal_nairobi}}

%\Figure

\subsection{Runtime}
The time measurements are made separately for each step of the quantum computing job execution process:
\begin{itemize}
    \item Creating
    \item Validating
    \item Queued
    \item Running
    \item Connection/communication
    \item Walltime
    \item Other
\end{itemize}

"Creating" and "validating" are the overhead to submit the job, compile and validate the quantum circuit. The "queued" time presents the time the job needs to wait for the hardware to become available. The "running" time represents the time spent on the quantum hardware or simulator. The "connection" time shows the time spent between the host sending the job and the framework running the job. The "walltime" is the time measured between starting the job and completing it on the host computer. The difference in time between "connection" and "walltime" is the time spent on the classical computer. Any time lost on other miscellaneous tasks is shown by "other".

\begin{figure}[h]
    \centering
    \includegraphics[width=0.9\linewidth]{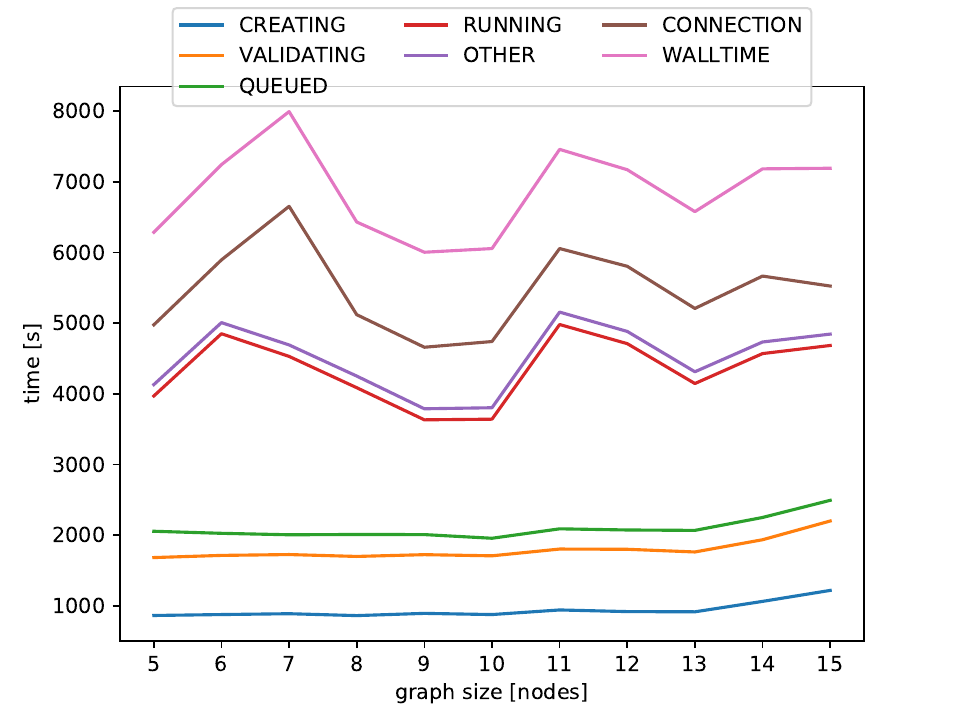}
    \caption{Results of the MCP benchmark for a problem size of 5 up to 15 nodes on the IBM QASM simulator.}
    \label{fig:QASM_mcp_15}
\end{figure}

This section presents the results of the MaxCut problem (MCP) benchmark. The results of the DSP and TSP benchmark are summarized. The times are shown cumulatively. The times are not normalized to the number of optimization steps. This is why the runtime for the simulator is significantly longer compared to the quantum hardware results, as the simulation has a higher limit on optimization steps. The results for the IBM QASM simulator are shown in Figure \ref{fig:QASM_mcp_15}. Since the simulator is relatively fast for these small problem sizes, a significant portion is spent on the classical computation.

%QASM

\begin{figure}[h]  
%\begin{subfigure}[b]{0.49\textwidth}
%    \includegraphics[width=\linewidth]{figures/mcp_ibmq_montreal_5_to_24.pdf}
%    \caption{Results of the MCP benchmark for 5 to 24 nodes on the IBM Montreal.}
%    \label{fig:montreal_mcp_24}
%  \end{subfigure}
%  \begin{subfigure}[b]{0.49\textwidth}
    \includegraphics[width=0.9\linewidth]{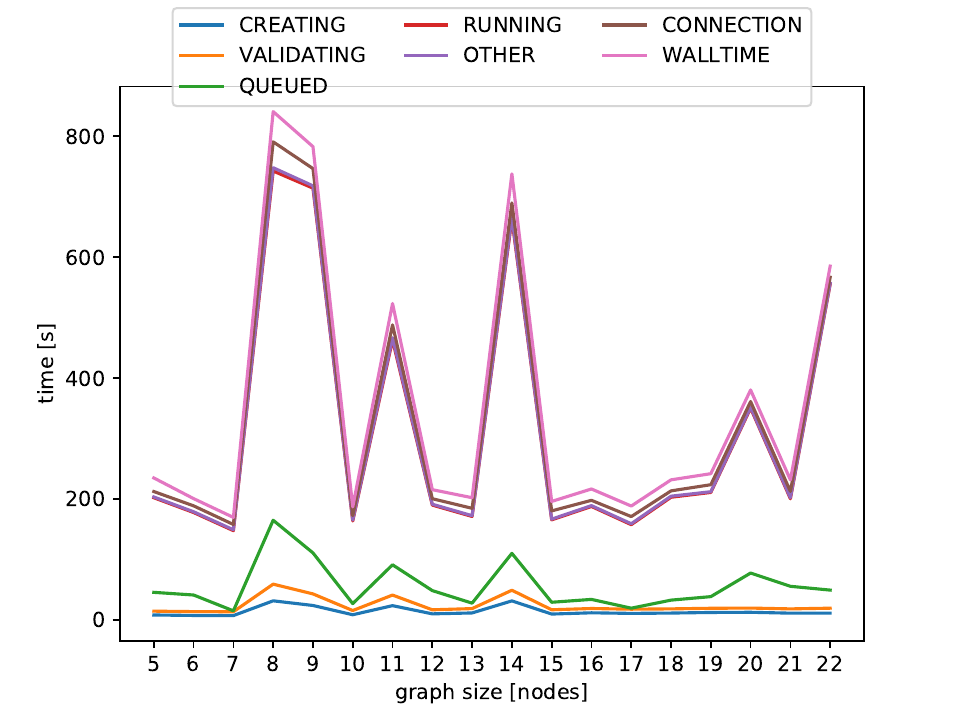}
    \caption{Results of the MCP benchmark for 5 to 22 nodes on the IBM Montreal.}
    \label{fig:montreal_mcp_22}
%  \end{subfigure}
%  %

%  \caption{Runtimes of the MCP benchmark on the IBM Montreal quantum computer}
\end{figure}

The results for the MCP benchmark on the IBM Montreal are shown in Figure \ref{fig:montreal_mcp_22}% and Figure \ref{fig:montreal_mcp_24}.
For Figure \ref{fig:montreal_mcp_22}, the results for problem size 23 and 24 were measured but have been cut out as these runtimes were dominated by queue times. These queue spikes measured at about 10K s clearly show the bottleneck for hybrid quantum computing without having a dedicated classical computer near the quantum computer or without exclusive access to a quantum computer. For most operations, the main part of the computation is running the quantum circuit as shown in Figure \ref{fig:montreal_mcp_22}. %However, as shown by the results in Figure \ref{fig:montreal_mcp_24}, the computation time can be completely dominated by the queuing time.

\begin{figure}[h]
    \includegraphics[width=0.8\columnwidth]{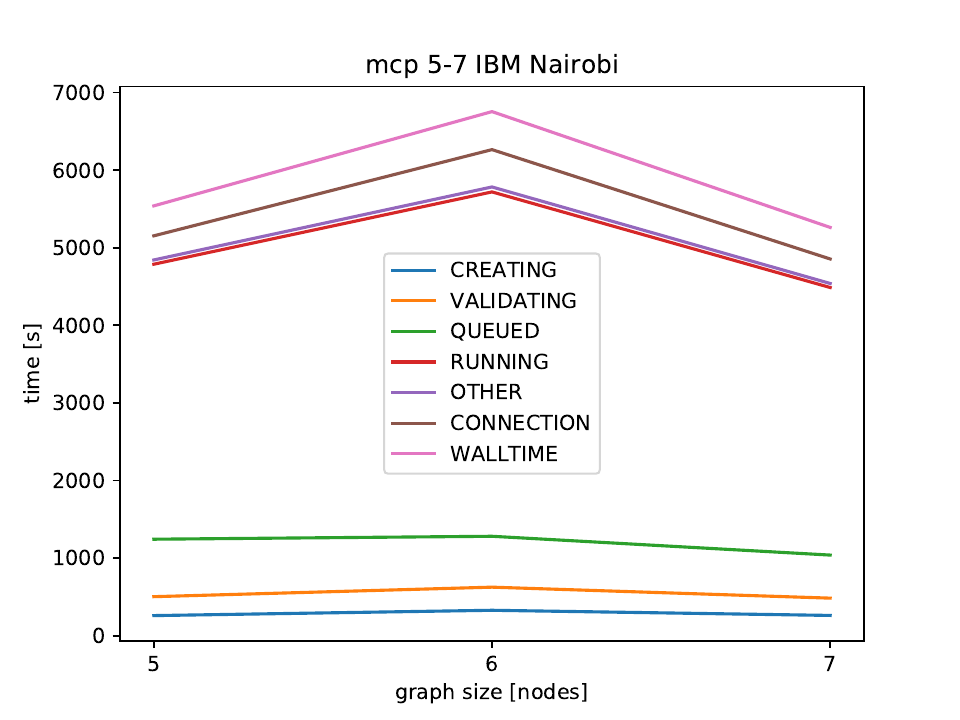}
    \caption{Results of the MCP benchmark up to 7 nodes on the IBM Nairobi.}
    \label{fig:nairobi}
  \end{figure}

\begin{figure}[h]
    \includegraphics[width=0.8\columnwidth]{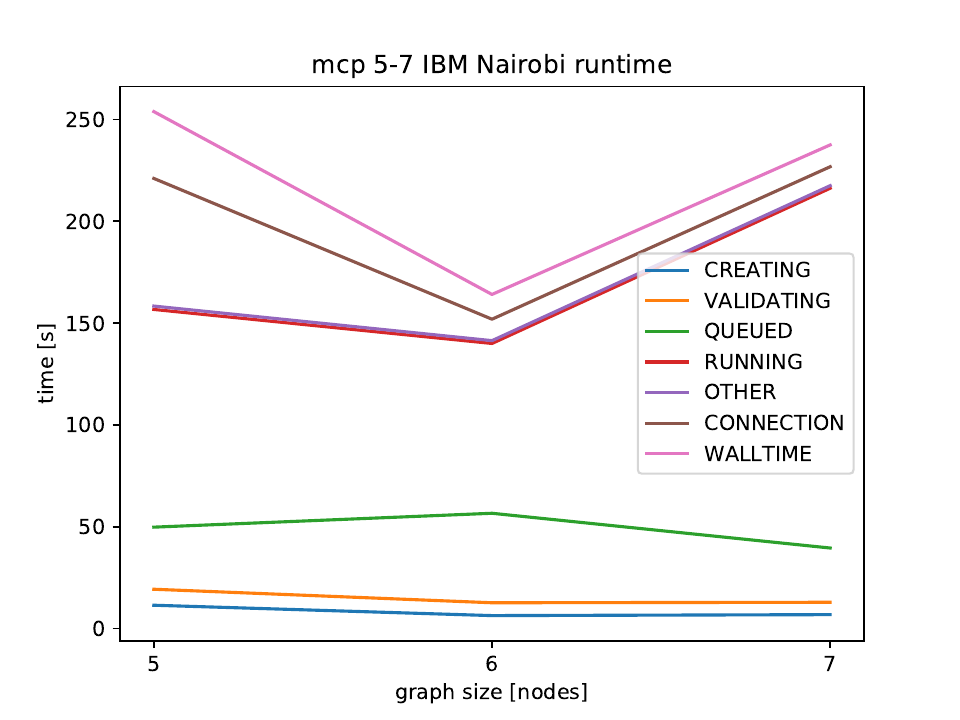}
    \caption{Results of the MCP benchmark up to 7 nodes on the IBM Nairobi with runtime environment.}
    \label{fig:nairobi_runtime}
  \end{figure}

The results for the MCP benchmark on the IBM Nairobi are presented in two parts: normal operation (Figure \ref{fig:nairobi}) and with the IBM runtime environment enabled (Figure \ref{fig:nairobi_runtime}).
The results of the IBM Nairobi can be compared to the Montreal hardware. The IBM Nairobi has been recently updated to support the runtime environment, and is expected to have increased in performance. The Nairobi hardware, however only support up to 7 qubits.
The IBM Nairobi using the runtime environment (Figure \ref{fig:nairobi_runtime}) unfortunately does not show significant speedups compared to the IBM Montreal (Figure \ref{fig:montreal_mcp_22}). By comparing the results of the Montreal hardware (Figure \ref{fig:montreal_mcp_22}) to the regular operation of the Nairobi hardware (Figure \ref{fig:nairobi}), it can be observed that the Nairobi hardware is significantly slower compared to the Montreal hardware. Possibly, due to the limited number of qubits, more \textit{SWAP} gates are required compared to the Montreal hardware. Alternatively, the quantum chip for the IBM Nairobi might be optimized for fidelity, in return for slower execution time. Only when implementing the IBM runtime environment (Figure \ref{fig:nairobi_runtime}), a performance similar to that of the Montreal hardware is achieved for graph size $5$ to $7$.
By comparing the runtime environment (Figure \ref{fig:nairobi_runtime}) to the regular operation on Nairobi (Figure \ref{fig:nairobi}), a significant speedup can be observed of 26.8 on average.

%Nairobi

%\begin{figure}[h]
%  \caption{Comparison for the runtime environment on the IBM Nairobi using the MCP benchmark.}
%  \label{fig:nairobi_mcp}
%\end{figure}

The runtime results of the DSP and TSP benchmark on the IBM Montreal and qasm simulator are summarized in Table \ref{tab:DSP_runtime}.

%DSP, TSP runtime
\begin{table}
 \begin{tabular}{|c|c|c|}
\hline
    DSP & & \\
    \hline
    nodes & Montreal & QASM simulator \\
    \hline
    5 & 27419.52 & 37.52 \\
    6 & 8483.87 & 49.95 \\
    7 & 4240.66 & 36.33 \\
    8 & 778.94 & 200.53 \\
    9 & 681.78 & 36.83 \\
    10 & 869.69 & 36.88 \\
    11 & 1119.32 & 37.32 \\
    12 & 1274.41 & 22.81 \\
    13 & 827.28 & 37.18 \\
    14 & 985.49 & 2547.42 \\
    15 & 1605.43 & 2706.53 \\
    16 & 948.95 & 464.76 \\
    17 & 982.52 & 505.48 \\
    18 & 1327.53 & 702.02 \\
    19 & 5070.27 & 742.48 \\
    20 & 4128.96 & 804.73 \\
    21 & 5034.03 & 1080.83 \\
    22 & 11464.99 & 1755.65 \\
    23 &  & 3671.98 \\
    \hline
    TSP & & \\
    \hline
    5 &  971.87 & 1314.93\\
    \hline
\end{tabular}
\caption{Measured total runtime for DSP and TSP benchmarks for the IBM qasm simulator and the IBM Montreal}
\label{tab:DSP_runtime}
\end{table}

\subsection{Outcome accuracy (score)}
This section presents the second part of our benchmark measurement called outcome accuracy (or score) achieved by the benchmark. The optimal scores are obtained by a brute force search. All measured solutions are obtained with parameter $p=1$. The score here refers to the cost of the MCP cost function i.e. the number of cut edges in a graph. This score is therefore calculated for edges $e_{jk}$ and ${-1, 1}$ representation of the measured state $\mathbf{z}$:
\begin{equation}
        c = \sum_{e} 0.5 * (1 - z_j * z_k)
\end{equation}

\begin{figure}[h]
    \centering
    \includegraphics[width=0.9\linewidth]{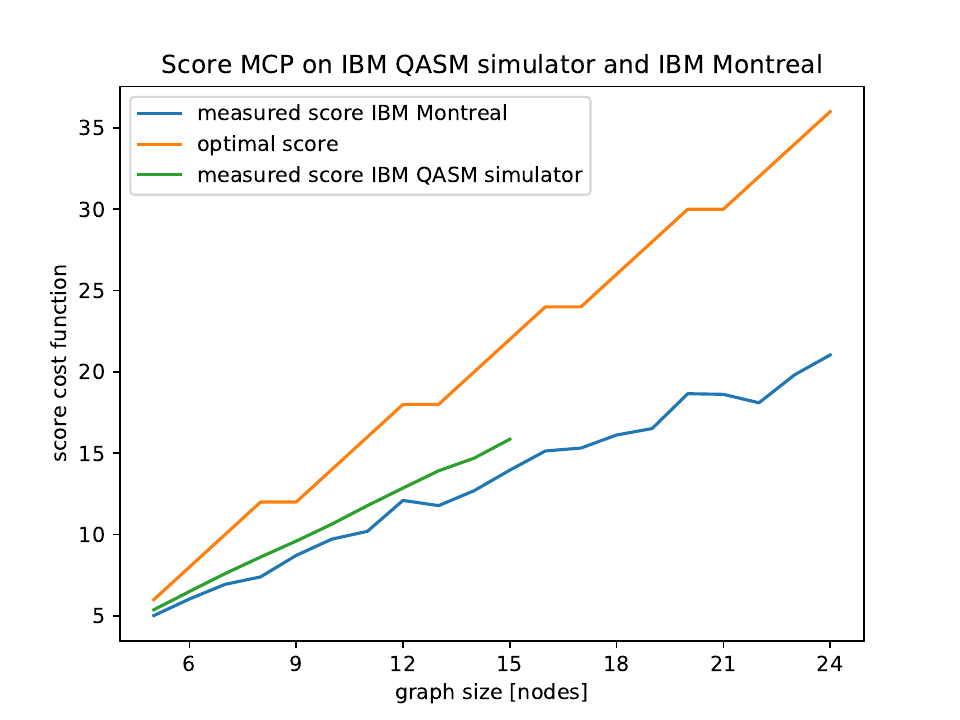}
    \caption{Comparison of the score results for the MCP benchmark on the IBM QASM simulator and the IBM Montreal.}
    \label{fig:score_QASM_montreal_mcp}
\end{figure}

As the problem size increases, so does the number of edges ( $nr\_edges = 2*nr\_nodes$ for the selected 4-regular circular graph). This means that in the MCP benchmark the number of cuts should rise as well. The plots of the optimal score show that this increase is almost linear. The scores presented are both the optimal scores for a given graph size (brute force search), as well as the measured score through the QAOA implementation. In Figure \ref{fig:score_QASM_montreal_mcp}, the measured MCP scores are given for the IBM QASM simulator for graph sizes 5 to 15. The score measured increases almost linearly with the graph size, as expected. The MCP scores for the IBM Montreal for graph sizes 5 to 25 are also presented in Figure \ref{fig:score_QASM_montreal_mcp}. Similarly to the IBM QASM simulator results, these measured scores increase almost linearly as well. In Figure \ref{fig:score_QASM_montreal_mcp}, the scores measured for the IBM QASM simulator and IBM Montreal are converted to the outcome accuracy.

In Table \ref{tab:mcp_score_nairobi}, the MCP score results for the IBM Nairobi are presented for graph sizes 5 to 7 as well as the scores for the IBM Nairobi, using the IBM runtime environment.

The MCP scores for the IBM Montreal, qasm simulator (Fig. \ref{fig:score_QASM_montreal_mcp}) and IBM Nairobi (Table \ref{tab:mcp_score_nairobi}) all show very similar results, though the measured score is predictably lower than the optimal score. The relatively low scores can be explained by the limit of $p=1$ and the limited amount of function evaluations on the classical optimizer. Increasing either will significantly increase runtime, but will show higher scores. This trade-off is further discussed in Section \ref{sec:scalability}.

\begin{figure}[h]
    \centering
    \includegraphics[width=0.9\linewidth]{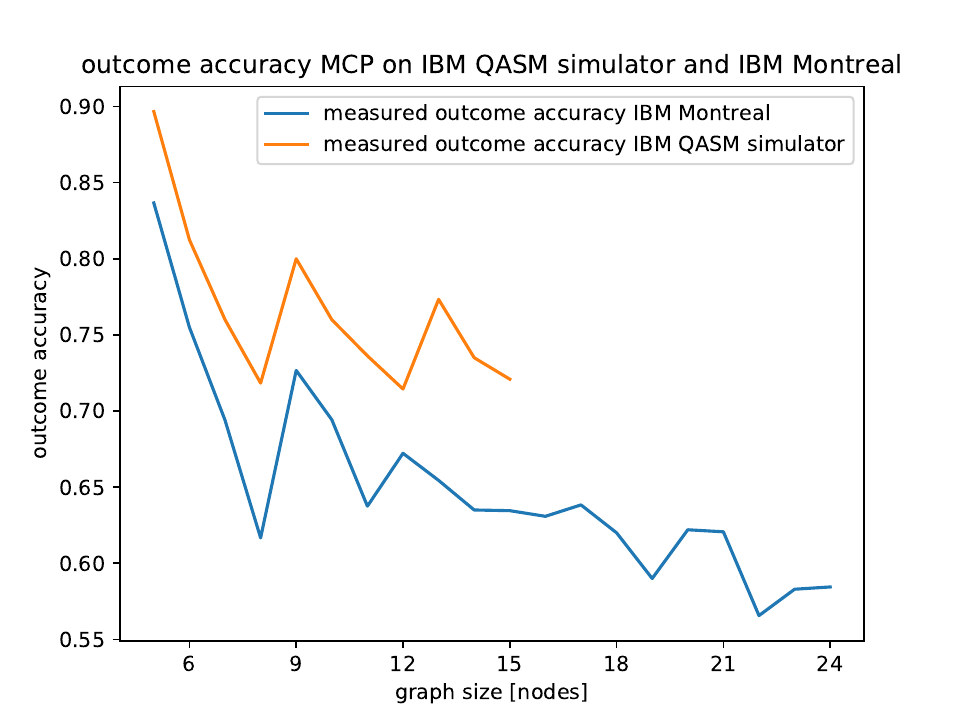}
    \caption{Comparison of the outcome accuracy results for the MCP benchmark on the QASM simulator and the IBM Montreal.}
    \label{fig:acc_QASM_Montreal_mcp}
\end{figure}

In the results of the outcome accuracy of both the IBM QASM simulator and the IBM Montreal (Figure \ref{fig:acc_QASM_Montreal_mcp}), an exponential decay of the outcome accuracy can be seen. Unsurprisingly, the outcome accuracy for the IBM QASM simulator lies higher than the outcome accuracy of the Montreal hardware. This is mainly due to noise, present in real quantum hardwares.

\begin{table}
 \begin{tabular}{|c|c|c|c|c|c|}
\hline
  & & \multicolumn{2}{|c|}{Nairobi}&   \multicolumn{2}{|c|}{Nairobi runtime}\\
    \hline
     nodes & max score &  score &  accuracy & score & accuracy \\
    \hline
    5 & 6 & 4.42 & 0.736 & 4.68 & 0.78 \\
    6 & 8 & 5.296 & 0.662 & 5.56 & 0.695 \\
    7 & 10 & 6.41 & 0.641 & 6.64 & 0.664\\
    \hline
\end{tabular}
\caption{Measured scores and accuracy for the MCP benchmark on the IBM Nairobi, with and without runtime environment enabled}
\label{tab:mcp_score_nairobi}
\end{table}

\iffalse
\begin{figure}[h]
    \centering
    \includegraphics[width=\linewidth]{figures/score_MCP_Nairobi_both.pdf}
    \caption{Comparison of the score results with and without the runtime environment on the IBM Nairobi.}
    \label{fig:score_nairobi}
\end{figure}
\fi

Despite the fact that the IBM Nairobi only supports up to 7 qubits, meaningful results are presented for the MCP benchmark for graph sizes of up to 7 nodes. While the scores for the regular run and the runtime accelerated implementation are very similar, the faster execution time on the runtime environment could enable larger $p$ configurations and therefore better scores for comparable execution time.
 
\iffalse
 \begin{figure}[h]
     \centering
    \includegraphics[width=\linewidth]{figures/accuracy_MCP_Nairobi_both.pdf}
     \caption{Comparison of the outcome accuracy results with and without the runtime environment on the IBM Nairobi.}
     \label{fig:acc_Nairobi_mcp}
 \end{figure}
\fi

The DSP and TSP benchmarks were run on both the IBM qasm simulator and the IBM Montreal. The results are summarized in Table \ref{tab:DSP_score}.

%DSP, TSP
\begin{table}
 \begin{tabular}{|c|c|c|c|c|c|}
\hline
  DSP & & \multicolumn{2}{|c|}{Qasm simulator}&   \multicolumn{2}{|c|}{IBM Montreal}\\
    \hline
     nodes & max score &  score &  accuracy & score & accuracy \\
    \hline
    5 & 6 & 7.22 & 0.802 & 7.08 & 0.787 \\
    6 & 8 & 8.83 & 0.883 & 8.86 & 0.886 \\
    7 & 10 & 10.11 & 0.842 & 10.12 & 0.843 \\
    8 & 12 & 11.72 & 0.837 & 11.71 & 0.836 \\
    9 & 12 & 13.43 & 0.839 & 13.28 & 0.83 \\
    10 & 14 & 14.66 & 0.814 & 15.11 & 0.839 \\
    11 & 16 & 15.94 & 0.839 & 16.7 & 0.879 \\
    12 & 18 & 17.62 & 0.839 & 17.69 & 0.842 \\
    13 & 18 & 18.87 & 0.820 & 18.76 & 0.816 \\
    14 & 20 & 20.63 & 0.825 & 20.6 & 0.824 \\
    15 & 22 & 21.88 & 0.810 & 22.71 & 0.841 \\
    16 & 24 & 23.18 & 0.827 & 23.05 & 0.823 \\
    17 & 24 & 24.77 & 0.826 & 23.38 & 0.779 \\
    18 & 26 & 26.06 & 0.814 & 25.18 & 0.786 \\
    19 & 28 & 27.57 & 0.811 & 27.32 & 0.8035 \\
    20 & 30 & 28.89 & 0.803 & 28.45 & 0.790 \\
    21 & 30 & 30.56 & 0.826 & 28.86 & 0.78 \\
    22 & 32 & 31.84 & 0.816 & 29.6 & 0.758 \\
    23 & 34 & 33.6 & 0.820 &  & \\
    \hline
    TSP & & & & & \\
    \hline
    5 & 200 & 269.7 & 0.651 & & \\
    \hline
\end{tabular}
\caption{Measured scores and accuracy for DSP and TSP benchmarks for the IBM qasm simulator and the IBM Montreal}
\label{tab:DSP_score}
\end{table}

\subsection{Scalability} \label{sec:scalability}
One of the main metrics discussed in Section \ref{metrics} is scalability. This includes how the runtime and the outcome accuracy scale for larger graph sizes. For the current quantum hardware, the challenge lies mainly in providing enough qubits. Especially in the context of QAOA, a system is desired to provide qubits as close as possible to the quantum supremacy threshold. This threshold was previously estimated to be in the few hundreds of qubits \cite{Guerreschi_2019}. The current largest system the QPack benchmark was tested on, is the IBM Montreal with 27 qubits. This is still far from the desired threshold.

As quantum hardware progressively becomes larger in terms of qubits and circuit depth, other aspects become more interesting. The scalability of the runtime will become one of the major metrics, but with the flexibility of QAOA, interesting trade offs arise. By increasing the number of QAOA layers $p$, the runtime becomes larger, but the outcome accuracy theoretically increases. The challenge then lies in optimizing the parameter to achieve the best results balanced for both metrics. Possibly, a focus can be placed on either of the metrics, in order to develop hardware specifically for fast runtimes or high accuracy. An interesting note on finding the optimal parameter $p$ for QAOA, is that increasing $p$ and therefore the quantum circuit depth, will also increase noise. This in turn will lower the outcome accuracy. This could mean that a different optimal $p$ could be found for each quantum hardware instance, which optimizes the outcome accuracy.

In the results shown in the previous section, a decrease of outcome accuracy is observed as the graph size increases. This is to be expected as the problems gets increasingly difficult for the algorithm to optimize. In order to adjust for this decrease in outcome accuracy, the general practice is to increase $p$. As previously discussed, this parameter is not optimized within the benchmark for multiple reasons, including the possible hardware dependency of this parameter.

\subsection{XACC results}

\begin{figure*}[h]
    \centering
    \includegraphics[width=0.6\linewidth]{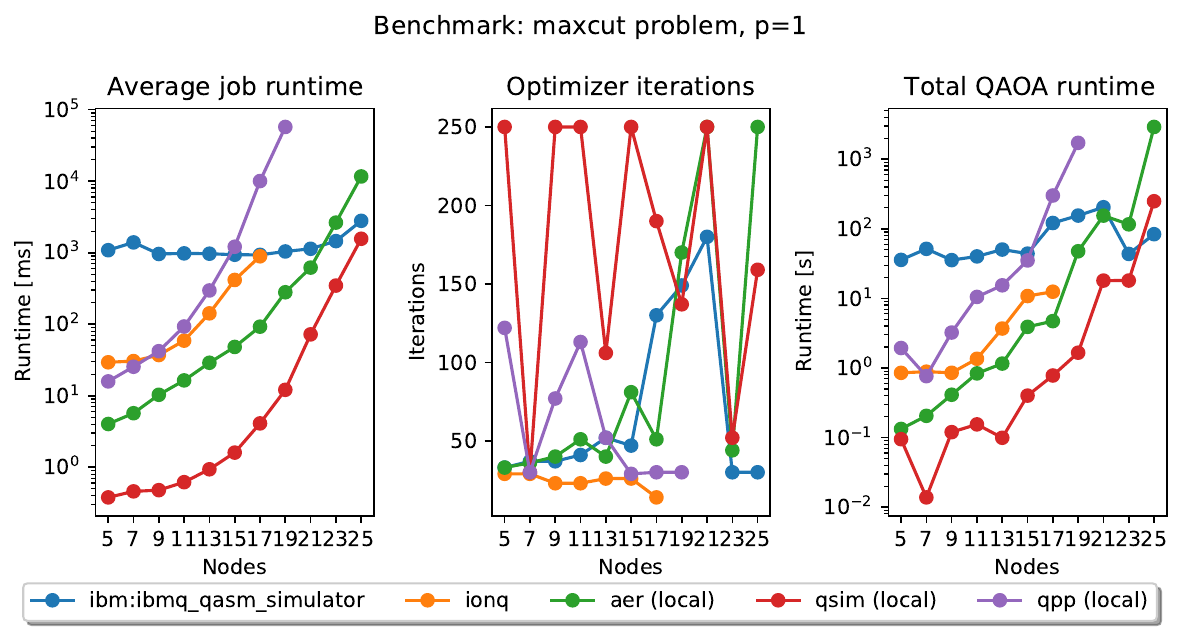}
    \caption{MCP benchmark runtime results for the QASM, IonQ, aer, qsim and qpp simulators for 5 to 25 nodes}
    \label{fig:xacc_maxcut}
\end{figure*}

This section covers the implementation of QPack using the XACC library \cite{XACC}, providing a proof-of-concept for a hardware-agnostic implementation of the benchmark. The XACC library allows for a single piece of code to run on different backends, as opposed to writing a new code for every backend provider. The aim of using this library is to create a universal benchmark: the user should not need to rewrite the benchmark in order for the benchmark to run on their system. This universal implementation shows how the QPack benchmark can be used to evaluate different quantum backends. Using the XACC library, the QPack benchmark was tested on 5 different simulators:
\begin{itemize}
    \item IBM QASM simulator (remote) \cite{qiskit}
    \item IonQ (remote) \cite{ionq_sim}
    \item aer (local) \cite{qiskit}
    \item qsim (local) \cite{qsim}
    \item qpp (local) \cite{gheorghiu2018quantum++}
\end{itemize}

The IBM QASM simulator is the same remote simulator as presented in previous sections. This simulator runs remotely through the IBM quantum experience. The second backend is the IonQ simulator, accessible remotely trough cloud services as well. The aer simulator is part of Qiskit and is run locally. The qsim simulator \cite{qsim} is part of Google's quantum AI project and run locally as well. The qpp simulator \cite{gheorghiu2018quantum++} is a C++ quantum computing library that is able to simulate 25 pure state qubits or 12 mixed state qubits. This simulator is again run locally. The local simulators aer, qpp and qsim are all integrated in the XACC library.

In Figure \ref{fig:xacc_maxcut}, the runtime results for the MaxCut problem for all mentioned simulators are presented. Results for DSP and TSP can be found in appendix \ref{sec:app_xacc}. The total runtime of the QAOA optimization (using COBYLA) is shown, as well as the average simulated quantum runtime for each optimizer iteration, as well as the number of iterations. All runtime results are presented on a logarithmic scale. The figure shows benchmark runs for each simulator, differing in the amount of maximally simulated qubits, 17 for IonQ, 19 for qpp and 25 for QASM, aer and qsim. Ideally, all backends would have been run for a problem size of 25 nodes, but the IonQ and qpp simulators showed difficulties simulating a larger number of qubits. All simulations were set to simulate 2048 shots. Note that because of execution and measurements on different platforms and the results showing just a single QAOA run, a fair comparison between simulators cannot be made. Nevertheless, these results can give some insight on the performance of different backends and provide information on what simulator may be favorable to use in certain applications.\\

From this figure can be seen that for most simulators the progression of both the average runtime and the number of iterations scale exponential.  One of the exceptions is that the average runtime of the IBM QASM simulator is much more constant (note that the logarithmic scale must be taken into account). The offset is however much larger. This means that for most of the 'smaller' graph sizes, the IBM QASM simulator is slower. This is only caused by the offset in simulated quantum run time, as the number of iterations is in the same order of size and has similar exponential progression as most of the other simulators. The logarithmic progression of the total simulated quantum runtime however suggests that the IBM QASM simulator should run faster for graph sizes much larger. The total runtime of the IBM QASM simulator already crosses the fastest increasing runtimes of the qpp simulator at graph size 15. The qsim simulator is however notably faster. The logarithmic progression of the IBM QASM simulator could be explained by the remote execution of the simulator, very likely providing much more computing resources. Interestingly, while the IonQ simulator is also run remotely, this simulator does not show the same linearity as the IBM QASM simulator, but increases exponentially, much like the local simulators. The offset of the simulated quantum runtime is still intriguing. One possible explanation could be that the IBM QASM simulator always simulates the maximum amount of qubits, also explaining the linearity.\\

The iterations of qpp and qsim show random progression. The number of iterations of the classical simulator remains unpredictable, as the algorithm might sometimes show much more difficulty finding the optimal points, compared to other similar runs, and vice versa. The number of iterations do not exceed 250, as a limit has been imposed to reduce outliers. The simulated runtime of qsim can be seen just overtaking the runtime of the IBM QASM simulator at 25 qubits. Due to a drop in iterations for graph sizes 23 and 25 for IBM QASM, the total runtime of qsim is just higher for these sizes. At approximately graph size 22, the IBM QASM simulator performs faster than the local aer simulator.

%% file: Conclusion.tex
\section{Conclusion and future developments}\label{sec:conclusion_future}

\noindent In this paper, the theoretical background on QAOA has been explored to set a basic understanding of how the algorithm operates and what the trade-offs are. Increasing accuracy by increasing $p$ requires longer runtime and is required for larger problems. The performance is also determined by the clause density and makes the performance problem dependent. State of the art development of the QAOA algorithm in practice is shown to be advancing, with a wide variety of algorithms shown for different, but similar problems. While these NP-hard problems are similar, it has been shown that QAOA is not well suited for every problem~\cite{sarkar2021estimating}. In some cases, such as the Ising problem, current classical approximations are expected to perform better and alternatives like the quantum adiabatic algorithm might be a better fit. For most of the problems discussed, the QAOA algorithm appears to be a good solution, with applications even reaching a full-stack application. The development of each implementation has been evaluated according to a TLR fitted for quantum algorithms. In this evaluation it is shown that most applications have at least been developed to the point that the algorithm has been simulated on a quantum simulator to show the capability of the implementations. In the case of the packing, MCP, DSP and TSP applications, a proper resource and performance analysis has been done. The implementation of TSP by Sarkar et al.~\cite{sarkar2020quaser} is currently the only implementation to have reached the level of a full stack development. This implementation also shows that current NISQ technology cannot properly support QAOA algorithms for practical implementation. Either advances on quantum hardware must be made, or more resilient implementations of QAOA must be developed.\\

Using this assessment, the MCP, DSP and TSP applications have been selected to be included in the QPack benchmark. Furthermore, a theoretical and numerical analysis has been presented on classical optimizers to find the implementation to best suit the QAOA algorithm. The results show that a clear trade-off exists between runtime and accuracy.\\

A reference implementation for QPack is presented with critical quantum benchmark requirements in mind \cite{blume2019metrics, resch2019benchmarking}. This benchmark will explore runtime and accuracy for available NISQ era quantum computers using various applications. The benchmark reflects the resources available, the implemented optimizations and the bottlenecks in the architectural implementation. The latter can specify the runtime on the quantum hardware, the classical hardware and the communication between both.\\

Results of the benchmark have been presented for several IBM systems and give insight to current bottlenecks and hardware improvements. A clear advantage was seen in the usage of the IBM runtime environment, using a remote classical computer. A significant speedup of $24 \times$ was observed, but expected to increase for larger problem instances. 

In future work, the benchmark can be further tailored for an improved implementation. The main pressing feature is to create a universal benchmark, using a quantum library. Libraries that currently explore this field are the LibKet \cite{LibKet} library and the XACC library \cite{XACC}.

For both libraries, the implementation in C++ will allow for better performance measurements for both runtime and memory consumption. The benchmark set presented by Qiskit uses the airspeed velocity (ASV) \cite{airspeed} tool to measure performance. This could be implemented to the current python implementation, but this effort is likely better spent on implementing such measurements on a C++ implementation. In the case of XACC, both could be done and a comparison between the python and C++ implementation could be presented.

An implementation using the XACC library is presented, showing progression towards a universal benchmark. With this implementation, a comparison is made between various simulators. The comparison highlights the advantages and disadvantages of the various simulators. While for example the QASM simulator performs much slower on smaller graph sizes, it outperforms other simulators at larger graph sizes ($\geq 13$). The qsim on the other hand excels at smaller graph sizes, but a much faster increase in runtime can be observed compared to other simulators. This shows that the benchmark is a viable tool for users to determine which simulator or hardware could be better tailored for their application.

Fine-tuning of the applications is also required, as a trade-off can be made between accuracy in runtime and precision when choosing the classical (hybrid) optimizer and the QAOA depth $p$. Finding the minimal requirements for the QAOA application to compete with classical approaches will need further examination. Currently, these parameters are left as user inputs.

A significant challenge in optimizing QAOA is parameter training. While studies have been performed using machine learning \cite{alam2020accelerating, Khairy2019ReinforcementLF, Wauters_2020}, other approaches have been presented as well. Several publications discuss the concentration of the $\beta$ and $\gamma$ parameters \cite{streif2019training, param_concentration, galda2021transferability}, which can be used to predict these parameters to speed up the parameter training. Unfortunately, for this benchmark implementation no similar results were observed. However, as multiple unaffiliated researchers have observed similar results, implementing this will certainly become a future improvement to the benchmark. 

Furthermore, the QAOA applications currently implemented are not optimal in terms of resources and can be further improved upon as mentioned in Section \ref{sec:app_measurements}. Finally, support for parallel processing is desired. As classical optimizers such as Nelder-Mead Simplex can be run in parallel \cite{ozaki2019accelerating, dennis1988parallel, lee2007parallel}, more objective functions need to run in parallel. If the quantum hardware can support a multiple of the qubits required by the objective function, such a parallel implementation can be supported. With current quantum hardware sizes, such implementations are not yet of importance, as practical problem sizes do not fit on the current hardware \cite{Guerreschi_2019}. As the quantum hardware scales further, this parallelism can result in significant speedup. \\

The QPack benchmark aims to expand its repertoire of algorithms in the future, as applications based on Shor's and Grover's algorithm \cite{sarkar2019algorithm} will become implementable on quantum computers. Currently, the quantum hardware cannot support these algorithms as more qubits and larger circuit depths are required. As more applications become practical on quantum computers, the QPack benchmark is envisioned to grow correspondingly to set the standard as a practical quantum algorithm benchmark. Interesting near term benchmark expansions include $\ket{0}$ bias measurements, the BB84 protocol and VQE for Ising problems.

\subsection{Code repositories} \label{sec:code_repos}
\noindent
The current implementation of the benchmark is available on the \textbf{\href{https://github.com/koenmesman/QPack}{github repository}}: https://github.com/koenmesman/QPack\\\\
The XACC implementation of QAOA benchmark can be found on this \textbf{\href{https://github.com/huub-d96/xacc_qaoa_benchmarks}{github repository}}: \\ https://github.com/huub-d96/xacc\_qaoa\_benchmarks.

%% file: QAOA_theory.tex
\section{Introduction to QAOA} \label{sec:details_qaoa}
\noindent As QAOA can be implemented to a larger variety of applications compared to VQE and has an advantage over QAA, QAOA was selected for the QPack benchmark. To understand this algorithm better for implementation, further details are discussed.\\

%The Quantum Approximation Optimization Algorithm (
QAOA is designed to find approximate solutions for combinatorial optimization problems with the help of quantum computing. To go in further detail, the algorithm will be explained using the equations given by Farhi et al. In combinatorial optimization problems, the goal is to maximize or minimize the number of clauses ($m$) satisfied. A clause ($\psi$) is a Boolean requirement, for example for a $n$-bit Boolean string  $\mathbf{z} = {z_1, ... z_n}$: 
\begin{equation}
    \psi_{ \rm example}: z_1 \land z_2
\end{equation}
To satisfy the clause of the example $z_1$ and $z_2$ must satisfy $z_1 = z_2 = 1$.

The QAOA algorithm depends on a parameter $p\geq 1$, which determines the quality of the approximation. Quality is given as an approximation ratio. This ratio is defined as either the number of clauses satisfied by the QAOA algorithm divided by the number of clauses satisfied in the optimal assignment:
\begin{equation}
\label{eq:opt_r}
    r=\frac{M_p}{C_{\max}(z)}
\end{equation}
or as the number of clauses satisfied by the QAOA algorithm divided by the number of clauses:
\begin{equation}
  r = \frac{M_p}{m}
\end{equation}
The latter is considered to be stronger \cite{lin2016performance}. The depth of the quantum circuit required to implement this algorithm grows linearly with $p$ times the number of constraints $m$ in the worst case \cite{farhi2014quantum}.

The equation that needs to be maximized (cost Hamiltonian) is as follows:
\begin{equation} \label{eq:max}
    C(z) = \sum_{\psi \in \Psi} C_{\psi}(z)
\end{equation}

Here, $C_{\psi}(z) = 1$ if clause $\psi \in \Psi$ is satisfied, and 0 otherwise. In the quantum approach we define the Boolean string as a vector $\ket{z}$ in the computational basis $\{\ket{0},\ket{1}\}$. A general guide to form such cost Hamiltonians from a cost function has been presented by Choi et al.\ \cite{design_hamilt}.
The operator for equation \eqref{eq:max} then becomes
\begin{equation}
    U(C, \gamma) = e^{-i\gamma C} = \prod_{\psi \in \Psi}e^{-i\gamma C_{\psi}}
\end{equation}
The terms of the product commute as these are diagonal in the computational basis and each term's locality is the locality of clause $\psi$. The unitary is dependent on angle vector $\bm{\gamma}$ and can be restricted to $ [0, 2\pi]$ since $C$ has integer values  \cite{farhi2014quantum}. Choosing the number of elements $\gamma$ is divided in, will impact precision and performance. To improve the precision of the QAOA algorithm, multiple cycles of the gates will be applied. The number of cycles is denoted as $p$. For integer $p \geq 1$, $\bm{\gamma}$ is defined as $\bm{\gamma} \equiv \{\gamma_1, ... ,\gamma_p\} $, where each $\gamma_i$ with $1 \leq i \leq p$ is an angle within the range $ [0, 2\pi]$.

To sum the outcomes of equation \eqref{eq:max}, the so-called mixer Hamiltonian $B$ or $H_B$ is introduced:\\
\begin{equation}
    B = \sum_{j=1}^{n}\sigma_j^x
\end{equation}
Here, the Pauli X-gates operate on a single qubit $\ket{z_j}$. For integer $p \geq 1$, $\bm{\beta}$ is defined as $\bm{\beta} \equiv \{\beta_1, ... ,\beta_p\} $, where each $\beta_i$ with $1 \leq i \leq p$ represents a angle $\beta$ for a single cycle. The general definition of the unitary applied to a single qubit then becomes:\\
\begin{equation} \label{eq:Bop}
    U(B, \beta) = e^{-i\beta B} = \prod_{j=1}^{n} e^{-i\beta \sigma_j^x}
\end{equation}
Here, $\sigma_j^x$ is the Pauli-X operator on qubit $j$.  The unitary $U(B, \bm{\beta})$ depends on angle vector $\bm{\beta}$, of which each $\beta_i \in \bm{\beta}$ runs from $0$ to $\pi$. The vector $\ket{z}$ is placed in superposition:
\begin{equation}
    \ket{s} = \frac{1}{\sqrt{2^n}}\sum_z\ket{z}
\end{equation}

The angle dependent quantum state can then be written as
\begin{equation}
    \ket{\bm{\gamma}, \bm{\beta}} = U(B, \beta_p)U(C, \gamma_p) ... U(B, \beta_1)U(C, \gamma_1)\ket{s}
\end{equation}
The expectation of $C$ then becomes:
\begin{equation} \label{eq:exp}
   F_p(\bm{\gamma}, \bm{\beta}) = \bra{\bm{\gamma}, \bm{\beta}}C\ket{\bm{\gamma}, \bm{\beta}}
\end{equation}
The maximum expectation for $p$ QAOA cycles ($M_p$) is then defined as:
\begin{equation}
    M_p = \max_{\bm{\gamma},\bm{\beta}}F_p(\bm{\gamma}, \bm{\beta})
\end{equation}
Since increasing parameter $p$ will increase the quality of the solution, in other words a equal or higher expectation will be found, we can state the following:
\begin{equation}
    M_p \geq M_{p-1}
\end{equation}
With the intent that $\lim_{p \to \infty}M_p = \max_z C(z)$, which is proven by Farhi et al.~\cite{farhi2014quantum}.

To give an intuition on how the algorithm operates, an example will be given. In this example, a system with 2 (qu)bits per clause is examined. The general clause will be worked out for bits $j$ and $k$:
\begin{equation}
    C = \sum_{<jk>}C_{<jk>}
\end{equation}
This is derived from equation \eqref{eq:max}. The expectation from equation \eqref{eq:exp} becomes:
\begin{equation}
\begin{split}
   & F_p(\bm{\gamma, \bm{\beta}}) =\\
   & \sum_{jk} \bra{s} U^{\dagger}(C, \gamma_1) ... U^{\dagger}(B, \beta_p)  C_{<jk>}U(B, \beta_p)  ...  U(C, \gamma_1)\ket{s}
\end{split}
\end{equation}
Here $U^{\dagger}$ denotes the conjugate transpose of the unitary operator $U$.

The operator for a single clause then becomes:

%%%%%%%%%%%%%%%%%%%%%%%%%%%%%%%%%
%general contribution:
\begin{equation}
    U^{\dagger}(C,\gamma_1)...U^{\dagger}(B,\beta_p)C_{<jk>}U(B,\beta_p)...U(C,\gamma_1)
\end{equation}

The contribution for $p=1$ can then be written as:
\begin{equation}
    U^{\dagger}(C,\gamma_1)U^{\dagger}(B,\beta_1)C_{<jk>}U(B,\beta_1)U(C,\gamma_1)
\end{equation}

Using an example for $C$, used in MaxCut applications (Appendix \ref{sec:MaxCut}), it can be shown which operators contribute to this equation. MaxCut and other applications will be discussed later in greater detail. For this example, the definition for the maximized equation becomes:

\begin{equation}
    C_{<jk>} = \frac{1}{2}(-\sigma^{z}_j \sigma^{z}_k +1)
\end{equation}
\\
Here, the Pauli-Z operator is denoted as $\sigma^z$. Now substituting equation \eqref{eq:Bop}, gives rise to the following:
\begin{equation}
    U^{\dagger}(C,\gamma_1)e^{i \beta_1 (\sigma^x_1+...+\sigma^x_n)}\frac{1}{2}(-\sigma^{z}_j \sigma^{z}_k +1)e^{-i \beta_1 (\sigma^x_1+...+\sigma^x_n)}U(C,\gamma_1)
\end{equation}

All $\sigma^z$ terms in the second exponent except for $\sigma^z_j$ and $\sigma^z_k$ can be moved in front of the $\frac{1}{2}(-\sigma^{z}_j \sigma^{z}_k +1)$ term as they do not interact with qubits $j$ and $k$, i.e. the operators commute ($\hat{A}\hat{B} = \hat{B}\hat{A}$).
Because these operators can be moved to the front, they cancel out with their Hermitian conjugate. For increasing $p$, $U(B,\beta)$ will contain the qubits connected one step further from the original clause. For example, if qubit $j$ shares a clause with qubit $l$, the operators acting on qubit $j$ will be included as well. By expanding this for increasing $p$, more qubits are considered for each clause, effectively increasing the quality of the approximation.
To expand this, the unitary $U(C,\gamma)$ must be understood. This unitary operator $C_{\alpha}$ is 1 if the condition is met and 0 otherwise. In the case of a graph (e.g. MaxCut), this means that the conditions are a connectivity matrix. This makes the term 1 if a connection between the qubits exists. The unitary of $\gamma_1$ will contain the $\sigma^z$ for j and k, but also all qubits which are connected. This has as consequence that $U(B,\beta_2)$  will keep $\sigma_j$ and $\sigma_k$ but also for all qubits connected to these qubits. This way, increasing $p$ will increase the qubits in the graph.

%% file: QAOA_examples.tex
\section{General optimization problems}
\label{sec:applications}
\noindent The general optimization problems discussed in this section will be the combinatorial satisfaction problems (CSP) max-$k$SAT and max-$k$XOR. Here, $k$ denotes the exact number of variables contained in a clause. The special cases of CSP will be discussed: the one with \textit{typicality} and the MaxCut problem. These problems represent general applications from which other practical applications can be derived. These practical applications are discussed in further detail in Appendix \ref{sec:selecting_app}, to determine which optimization problem solutions are most developed and suited for the benchmark.

%preparation qubits
%superposition using H gates
%quantum gate for objective function?

%Special cases
%   typicality
%   no overlapping results
%   max-cut

\subsection{Max-\texorpdfstring{$k$}{TEXT}SAT and max-\texorpdfstring{$k$}{TEXT}XOR}
\noindent In max-$k$SAT the objective function needs to be maximized, as discussed in section \ref{sec:Quantum_Alg}. Max-$k$SAT is the general set of problems where the number of Boolean constraints satisfied is maximized. Individual constraints can have their own weight. Giving a weight to constraints can be beneficial for certain problems. For now the unweighted problems are examined, i.e. $weight = 1$. There exist problems where not all constraints can be satisfied. In these cases the maximum number of constraints, but not all need to be satisfied. The problem can take form of any constraint that applies to $k$ parameters. Currently, the most common max-$k$SAT problem discussed, is the max-$k$XOR. For this variant of max-$k$SAT problems, the objective function is defined as \cite{lin2016performance}:
\begin{equation}
    C = \frac{1}{2}\pm \frac{1}{2}\prod_{i=1}^k x_i
\end{equation}

\subsubsection{MaxCut}\label{sec:MaxCut}

\begin{figure}[h]
    \centering
    \includegraphics[width=0.4\linewidth]{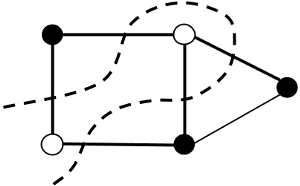}
    \caption{A MaxCut problem visualized as a graph \cite{max_image}}
    \label{fig:max}
\end{figure}

\noindent This is by far the most discussed problem with regard to QAOA \cite{willsch2020benchmarking, Guerreschi_2019, farhi2014quantum, Zhou_2020, crooks2018performance, alam2020accelerating, lin2016performance}.
The MaxCut problem is a max-2XOR problem, that can be visually represented as a graph. The individual parameters are presented as vertices, while the clauses can be represented as edges. The objective of MaxCut is to find as may edges $<jk>$as possible, where $j = (1-k)$. A visualization is shown in Figure \ref{fig:max}.\\ 

\begin{figure}
    \centering
    \includegraphics[width=0.8\linewidth]{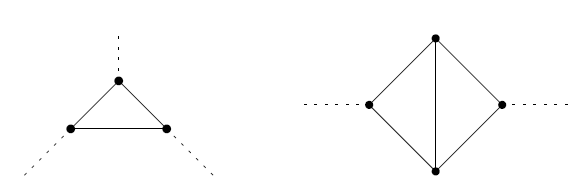}
    \caption{Possible sub-graphs for 3-regular graphs \cite{farhi2014quantum}}
    \label{fig:3-reg}
\end{figure}

MaxCut is a known NP-complete problem and therefore serves as a good example of what can be accomplished with QAOA \cite{conf/coco/Karp72, gary1979computers}. The performance of MaxCut can be increased with pre-processing. If the input graph has a regular shape, e.g. with a bounded degree, the graph can be divided into sub-graphs that occur multiple times. This way, only the unique graphs need to be analyzed, and their expectation value multiplied with their corresponding occurrence \cite{farhi2014quantum}. 
An example is the 3-regular graph, which only has 2 possible sub-graphs, as shown in Figure \ref{fig:3-reg}.
With pre-processing, only the occurrence of the different sub-graphs need to be found, and the QAOA needs to find the expectation value of the 2 possible graphs. This however implies that the classic algorithm needs to find the occurrences of the sub-graphs with high performance. How this carries over to practical applications is not yet explored, and will be very application dependent. A mayor concern is that finding the subgraphs, known as the clique problem, is in itself one of Karp's NP-complete problems \cite{conf/coco/Karp72}.
%why max cut -> practical future

MaxCut can be applied to challenges as circuit layout design \cite{maxcut_circuit}, data clustering \cite{maxcut_cluster}, network design and statistical physics \cite{WANG2010240}. This shows that MaxCut can be applied in a large variety of industry level optimization problems.
%quantum circuit design

\subsubsection{Typicality}
A CSP is considered typical if \cite{lin2016performance}
\begin{equation}
    \mathbb{E}_{\psi} [\psi(x)-\hat{\psi}(\emptyset)]\equiv \sum_i P(\psi_i)(\psi_i(x)-\hat{\psi}_i(\emptyset)) = 0,\quad \psi_i \in \Psi %z==>\alpha/psi
\end{equation}
where $P$ denotes the probability distribution for a constraint, with $\sum_i P(\psi_i,l)=1$ where $l$ is the $l$th constraint.
Applying the Fourier transformation on $\psi(x)$ gives:
\begin{equation}
    \mathbb{E}_{\psi}[\hat{\psi}(K)]=0 \quad \forall K \neq \emptyset
\end{equation}
This means that the Fourier coefficients associated with $P(\psi)$ have $0$ mean. Note that the MaxCut problem does not have typicality.
Typicality is interesting in CSP, as this leads to analytical derivation of the satisfaction ratio for this set of problems. Lin et al. have proven that for every CSP with a bounded degree and typicality, a satisfaction ratio of $\mu + \Omega (1/\sqrt{D})$ can be found with a probability of $1-O(D^3/m)$ \cite{lin2016performance}. Here the maximum number of constrains in which a variable occurs is $D$. The number of variables per constraint is constant, as the CSP has a bounded degree. With this probability, a high success rate can be achieved for $m\gg D$.

%% file: Selecting_app.tex
\section{Overview of NP-Hard problems with QAOA implementations} \label{sec:selecting_app}

\subsection{MaxCut}
The details of the MaxCut problem have been discussed in detail in Appendix \ref{sec:MaxCut}. The optimization can be applied for circuit layout design \cite{Barahona_circuit}, Data clustering \cite{maxcut_cluster} or implementing the Ising model. The MaxCut problem for QAOA has been studied  extensively \cite{crooks2018performance, willsch2020benchmarking, Zhou_2020, lin2016performance}. It has been shown that the approximation of QAOA achieves better results than the classical Goemans-Williamson \cite{Goemans} for $p \geq 8$ \cite{crooks2018performance}. For the general MaxCut algorithm, the gates required are $O(N^2P)$ and can be run in $O(NP)$ assuming $O(N)$ gates can be run in parallel. The classical Goemans-Williamson algorithm, in contrasts, requires a run time of $O(Nm)$ for $m$ edges \cite{crooks2018performance}.

\subsection{Ising model}
The Ising model can be approximated by the MaxCut algorithm \cite{jerrum1993polynomial}. Implementations for the Ising model can be found in physics for simulation of e.g. phase separation, lattice-based liquid-gas model or spin glasses. Applications for this model can also be found in biology for the protein folding problem \cite{protein}.  The model is also supported in Qiskit \cite{qiskit}, alongside the popular MaxCut algorithm. The Ising model has in recent publishing been compared for both classical and QAOA algorithms \cite{philathong2020computational}. It becomes apparent that for large clause density, QAOA shows reachability deficits \cite{Akshay_2020} and is therefore no improvement on the classical algorithms. For a large enough search space, the QAOA algorithm can reach a Grover scaling of $O(\sqrt{N})$.

%QUBO

%A[m x n].x[n]<=b[m] for m constraints, n variables/components
%reduce to clique partitioning: for graph (V,E) with weighted E, find the subset of V that forms a connected clique with the maximum/minimum weight.
\subsection{Set packing}
\begin{figure}
    \centering
    \includegraphics[width=0.9\linewidth]{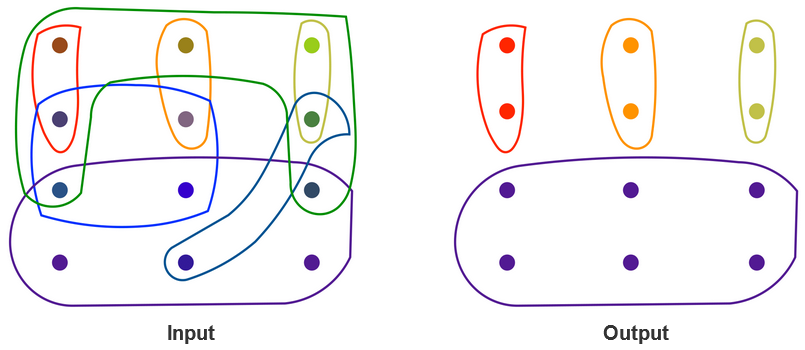}
    \caption{Visualization of the set packing problem \cite{setpacking}}
    \label{fig:set_pack}
\end{figure}

In the set packing problem a set of subsets is given. A minimal selection of the subsets must be chosen to cover all elements of the set \cite{setpacking}. A visual representation of this problem is shown in Figure \ref{fig:set_pack}.

This problem is also covered by Ruan et al. by adjusting the Hamming distance to $d(\mathbf{x}, \mathbf{x'}) = 1$. The complexity achieved with their method is $O(n)$ for $n$ subsets.
%applications

\subsection{Vertex cover}
In the vertex cover, all edges of a graph ($V,E$) must be covered by selecting the minimum amount of nodes. This problem is implemented by Ruan et al. with an alteration on the previous method \cite{ruan2020quantum}. The vertex cover problem has implementations in, e.g., monitoring. Examples are security camera placement or network link monitoring.

\subsection{Dominating set problem}
The dominating set problem is very similar to the vertex covering problem, but instead of covering edges, there must be an edge to each vertex from a vertex from the selected set. The applications of both problems are very similar as well. An implementation of QAOA and an analysis of the algorithm is presented by Nicholas Guerrero \cite{guerrero2020solving}. In the circuit presented, for $n$ vertices the following is required \cite{guerrero2020solving}:
\begin{itemize}
    \item $18n^2 - 6n$ single qubit gates
    \item $16n^2 - 12n$ controlled Pauli-X gates
    \item $2n$ qubits (of which $n$ are ancillary)
\end{itemize}

\subsection{Traveling salesman problem}
The traveling salesman problem is a well known problem where a salesman needs to visit all cities (vertices) which are connected by roads (weighted edges). A special case of the problem, the undirected Hamiltonian circuit, is also one of Karps's 21 problems. A (naive) classical approach would check all routes and would have a complexity of $\bf{O}(n!)$ \cite{TSP_blog}. An implementation based on Grover's algorithm has shown a quadratic speedup to $\bf{O}(\sqrt(n!))$ \cite{TSP_blog}, but such an implementation would require a lot of quantum resources and would currently not be applicable to quantum hardware. A QAOA solution has been presented by Radzihovsky et al. \cite{Radzihovsky2019AQS} and by Henry and Dudas \cite{hentry_dudas}. The respective source codes for implementation on a Rigetti (pyquil and the Rigetti QVM) quantum computer are available at \textbf{\texttt{\url{https://github.com/murphyjm/cs269q_radzihovsky_murphy_swofford}}} \cite{Radzihovsky2019AQS} and \textbf{\texttt{\url{https://github.com/danielhenry1/QAOA_TSP}}} \cite{hentry_dudas}. Radzihovsky et al. show that the number of gates in the cost Hamiltonian scales with $n^2$ for $n$ cities, and a circuit depth of $2n$ for even $n$ and $2(n-1)$ for uneven $n$. Ruen et al. also published an implementation using $n^2$ qubits. The traveling salesman problem has a large range of applications, such as computer wiring, drilling PCB and order-picking in warehouses \cite{Matai10}. A recent implementation of QAOA on the traveling salesman problem is presented by Sarkar et al. \cite{sarkar2020quaser}. In this implementation, the QAOA is applied for quantum accelerated de Novo DNA sequence reconstruction. An application and QAOA simulation are presented, but does not feature a hardware implementation as the currently available quantum hardware are not sufficient to support meaningful results \cite{sarkar2020quaser}.

\subsection{Tail assignment problem}
The tail assignment problem deals with the scheduling of a set of flights to a set of aircraft to create a feasible flight schedule, while keeping the lowest cost. This problem is an optimization of a packing problem with additional constraints. Vikstal et al. propose an implementation of QAOA for this problem \cite{vikstaal2020applying}. In their publication, they show that for p=2, 99.9\% can be achieved, but requires 74 measurements for 25 route instances. This case performs better than quantum annealing. Using more measurements instead of increasing $p$ is computationally less expensive.

\subsection{Facility placement problem}
The facility placement problem aims to place a node (facility) with the least transportation cost (this in itself is a Weber problem \cite{weber}), while considering other constraints. A QAOA implementation is presented by Quiñones and Junqueira \cite{quinones2020tailored}. This paper, however, shows that the QAOA implementation has a much higher cost and lower optimal solution probability than a VQE for the investigated case. This might indicate that this type of problem is not suited for QAOA.

\subsection{QAOA for Grover's algorithm}
An interesting approach has been presented in \cite{Jiang_2017}, where the principals of QAOA have been applied to Grover's algorithm. Much like the implementations discussed before, it uses $p$ iterations to improve fidelity/accuracy. This implementation could potentially make the algorithm that is proven to perform better than classical algorithms (quadratic speed-up), executable on relatively few qubits. This would allow it to run on near-future devices. Some critical issues however exist with this implementation. QAOA has limited accuracy, while Grover's algorithm can only be applied practically (e.g. database searching) if an exact solution can be found. The authors mitigate this problem by checking with multiple rounds whether the found solution is correct. This might require the algorithm to run for many iterations before the solution is found. Furthermore, the chance of success of the algorithm is only 50\%, making it impractical for implementation. Error-correction schemes might make this implementation worthwhile, but it is unclear whether such a solution can be found. Quantum error correction in adiabatic models is still a challenge \cite{Young_2013}, but could make this application implementable.

%classic algorithms: mixed integer linear programming

%% file: Parameter_optimization.tex
\section{Parameter optimization}
\label{sec:param_opt}

This section explores the strategies to optimally determine the $\beta$ and $\gamma$ parameters. The optimization of these parameters is considered to be the main bottleneck of QAOA \cite{sarkar2020quaser}, and the selection of the optimizer could make or break the application. This single objective real-parameter optimization is done classically and many algorithms have been developed for this problem. Critical parameters for choosing the classical optimization are:

\begin{itemize}
    \item Number of parameters
    \item Problem size
    \item Smoothness of the objective function
    \item Local optima
\end{itemize}

\subsection{Gradient-based optimization}
Gradient-based optimization uses the gradient of the problem function to find a (local) optimal. Gradient-based optimizers converge generally faster than gradient-free ones and would therefore be favorable \cite{opt_ch3}. However, gradient-based optimization only works well on smooth functions and is prone to getting stuck in local optima. Due to the dependency on smooth functions, any function that is either discontinuous or noisy will cause problems for the optimization. In the case that a smooth function is available, the problem of local minima could be averted by choosing multiple starting points and running the optimization multiple times or in parallel if possible. This, however, increases the computational complexity and choosing an efficient algorithm for the starting value is not trivial. Often random points are chosen, but a more efficient algorithm based on the Discrete Fourier Transform (DFT) is presented by \cite{Zhou_2020} but is not open-source.

%The explored Gradient-Based algorithms are:
%\begin{enumerate}[label*=\arabic*.]
%    \item Newton method
%    \item Quasi-Newton method
%    \begin{enumerate}[label*=\arabic*.]
%        \item BFGS
%        \item square root BFGS
%        \item DFP update
%        \item limited memory BFGS
%        \item non-linear Conjugate Gradient
%    \end{enumerate}
%    \item Gradient Descent
%    \begin{enumerate}[label*=\arabic*.]
%        \item Batch
%        \item Stochastic
%        \item Mini-batch
%    \end{enumerate}
%    \item Steepest Descent
%    \item Trust region
%\end{enumerate}

\subsection{Gradient-free algorithms}
Gradient-free algorithms do not depend on a gradient of the function, as the name suggests. This means that these algorithms can be applied to objective functions that are noisy, discrete or discontinuous \cite{opt_ch6}. Gradient-free can be more resilient with respect to local minima, depending on the algorithm. The downside of these algorithms is that these generally take more computation time or more iterations.

\subsection{Problem instances}
The objective function is evaluated with the above mentioned critical aspects, in order to find the optimization algorithm that fits best. The objective functions are considered in general and not per individual problem. The reason for this will become apparent in the evaluation.

\textbf{Number of parameters}: For all problems, the number of parameters are $2\times p$: $\beta$ and $\gamma$ for each QAOA iteration $p$. Optionally, $p$ can be taken as a parameter as well, but since publications show that QAOA is most effective in low depth, i.e. low $p$, this parameter will not contribute significantly. Generally a value for $p$ is chosen beforehand (e.g. Crooks \cite{crooks2018performance} shows that $p \geq 8$ is required to outperform the Goemans-Williamson algorithm) after which $\beta$ and $\gamma$ are optimized. Theoretically, an optimizer would increase $p$ to infinite as this would give the best results. In practice, $p$ can optionally be included as well, as performance is likely to decrease with higher circuit dept on quantum hardware.

\textbf{Problem size}: The problem size is expected to grow very large, as scalability needs to be considered. This does not differ per problem type.

\textbf{Smoothness of objective function}: The smoothness of the objective function is expected to differ per problem type. However, due to the probabilistic nature of the results, as well as noise in quantum hardware, the objective function is never smooth in practice.

\textbf{Local optima}: The presence of local optima is dependent on each individual problem, but due to noise in quantum hardware, local optima might still occur. Local optima might not be as strong as local minima inherent to the objective function. This could give room to optimizers that have some trouble escaping from strong local optima.

\subsection{Selection of optimizer} \label{sec:c_opt}
\noindent Considering how all QAOA problem instances will suffer from noise, discontinuities and local optima, no gradient-based optimizer can be applied. From the gradient-free algorithms the following are considered:
\\

\textbf{Nelder-Mead Simplex}: Nelder-mead is efficient for non-smooth objective functions. It is however not very efficient for 10 or more design variables. Considering that only $2\times p$ variables are optimized, the Nelder-Mead optimizer fits well for low depth QAOA. Nelder-Mead is also more resilient with regard to local optima compared to other local optimizers.\\

\textbf{Genetic Algorithms}: Genetic algorithms generally require large number of function evaluations, but is applicable to multi-objective optimizations. Genetic algorithms are not well suited, since QAOA problems only have one objective function. The large number of iterations are therefore inefficient.\\

\textbf{Simulated Annealing (SA)}: Simulated annealing has more freedom than other gradient-free optimization algorithms, but comparison to other algorithms shows that the computation per iteration of SA is less efficient. Alternatively, SA could be accelerated with VQE, but the performance is not yet adequately explored and will be more demanding of the quantum hardware. Potentially, this could be used in the future when quantum hardware is more mature and scalable.\\

\textbf{Particle swarm optimization (PSO)}: Particle swarm optimization can converge fast, has a short computational time and has very few parameters to adjust. The downsize is that it is prone to get trapped in local optima, especially with complex problems \cite{optmization_PSO}. Since large problem instances need to be considered with regards to scalability, PSO might not be suited for QAOA. PSO could be considered for problem instances with no inherent local optima (i.e. local optima due to noise). Unfortunately, it could be difficult to tune initial parameters for PSO. Considering that QAOA should be applied to multiple different problems, tuning the parameters could create a bottleneck. PSO could be considered an option for similar problems and a custom PSO parameterization. A custom PSO could be considered for the Ising model, with seemingly no inherent local optima \cite{pagano2020quantum}, but not for the more popular and more developed MaxCut  \cite{Khairy2019ReinforcementLF}.\\

\textbf{Symmetric Rank-1 Update Method (SR1)}: The SR1 method is not a gradient-free algorithm, but is included for special cases. SR1 uses a first-order derivative, as opposed to a second-order derivative like the Newton method. This means that the smoothness requirement on the objective function is less strict. In the case that the hardware has sufficiently low noise and multiple measurements per iteration are performed, the SP1 method might be applicable. The main difficulty, is that it is subjective to how smooth an objective function needs to be for SP1 to perform well, as no clear comparison is done between noise levels in objective functions and performance of the algorithm. It would also require the objective function to be evaluated, meaning it would be unclear per implementation if the algorithm works well. Quasi-Newton methods such as SP1 and BFGS only work for finding local minima. This means that if multiple local minima exist in the solution landscape, the global minima might not be found. With this in mind, SR1 can only be applied to objective functions with only one minimum and would require extensive evaluation of the objective function. Another Quasi-Newton method that is commonly used is the BFGS algorithm, but as long as the problem field has a limit constraint (in this case $\gamma, \beta \in [0, 2\pi]$), SR1 converges faster. \\

More state-of-the-art algorithms exist, as competitions in the field of single objective real-parameters optimization are organized by IEEE yearly (CEC Real-Parameter Optimization Competitions) but since the winning algorithms have not been found as either open-source or a library, the winning algorithms are not considered for application. Regardless, the best algorithm for low dimensions was found to be UMOEA \cite{CEC}.\\

From the evaluated optimizers, Nelder-Mead is considered the best fit for QAOA. Numerical comparison of classical optimizers \cite{num_eval} show that in terms of performance, BFGS and Nelder-Mead differ very little in runtime and find identical results in sufficiently smooth functions. This raises the argument that if the objective function is sufficiently smooth, Nelder-Mead could be used as well and no custom optimizer needs to be applied. This saves development time for the application. PSO or SR1 could be considered only for very special cases. For the general case, UMOEA could be implemented to possibly further improve performance.

%\cite{Zhou_2020}: NM slightly outperforms BFGS, even with initial point generation
%EDIT: NM is NOT global, but generally performs better with few local minima

%% file: results_appendix.tex
\section{DSP and TSP benchmark results}
\label{app:results-qiskit}
\subsection{DSP}
\begin{figure}[h]
    \includegraphics[width=0.4\textwidth]{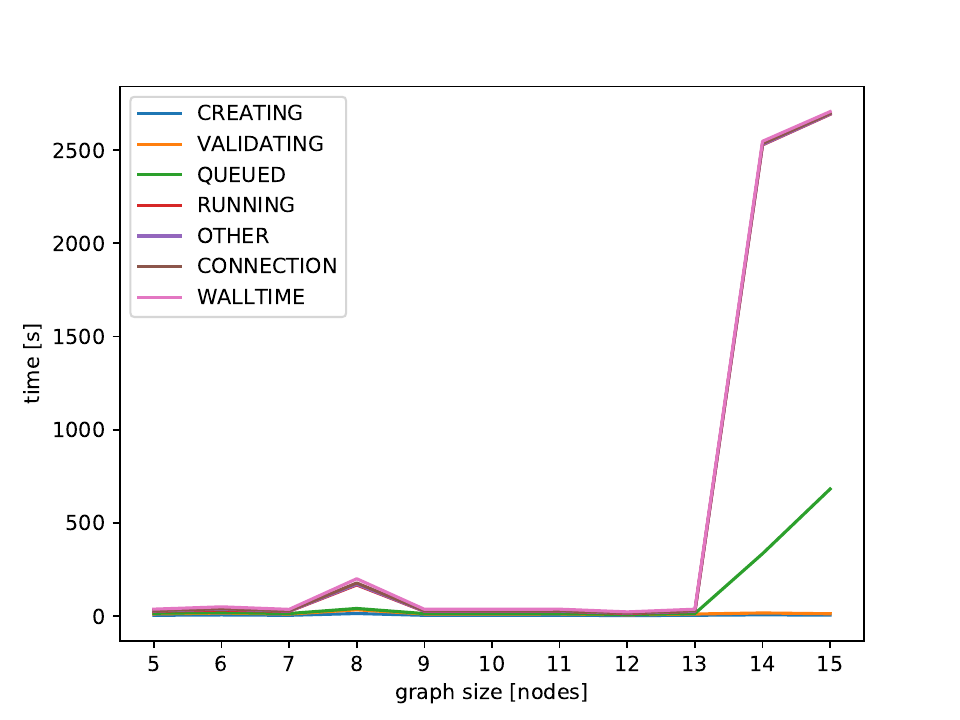}
    \caption{Results of the DSP benchmark up to 15 nodes on the IBM QASM simulator.}
    \label{fig:QASM_dsp_15}
  \end{figure}

\begin{figure}[h]
    \includegraphics[width=0.4\textwidth]{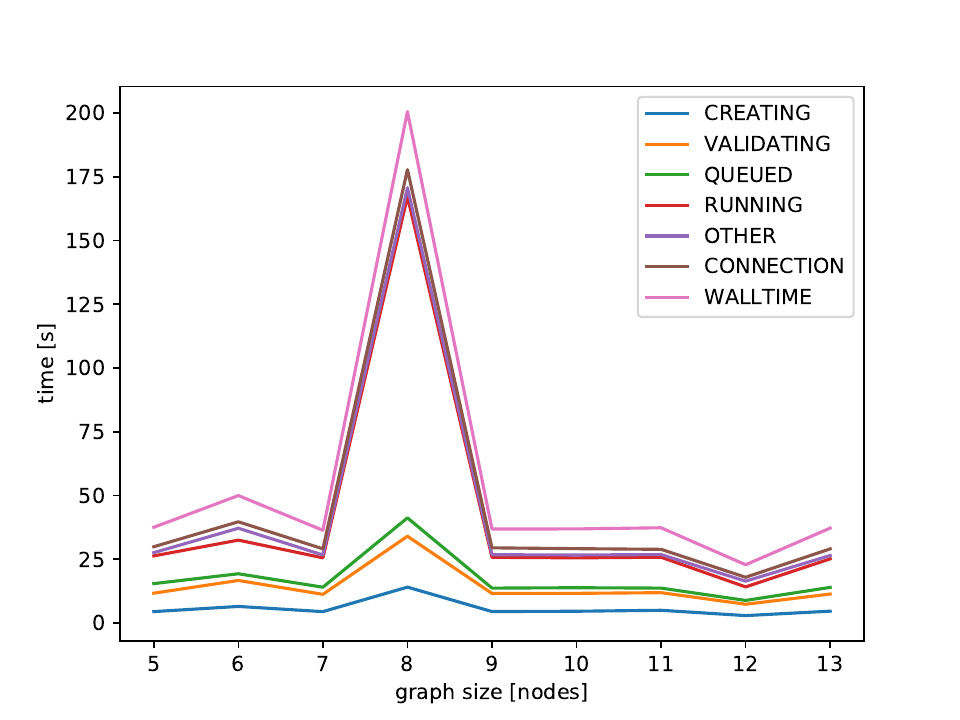}
    \caption{Results of the DSP benchmark up to 13 nodes on the IBM QASM simulator.}
    \label{fig:QASM_dsp_13}
  \end{figure}

\begin{figure}[h]
    \includegraphics[width=0.4\textwidth]{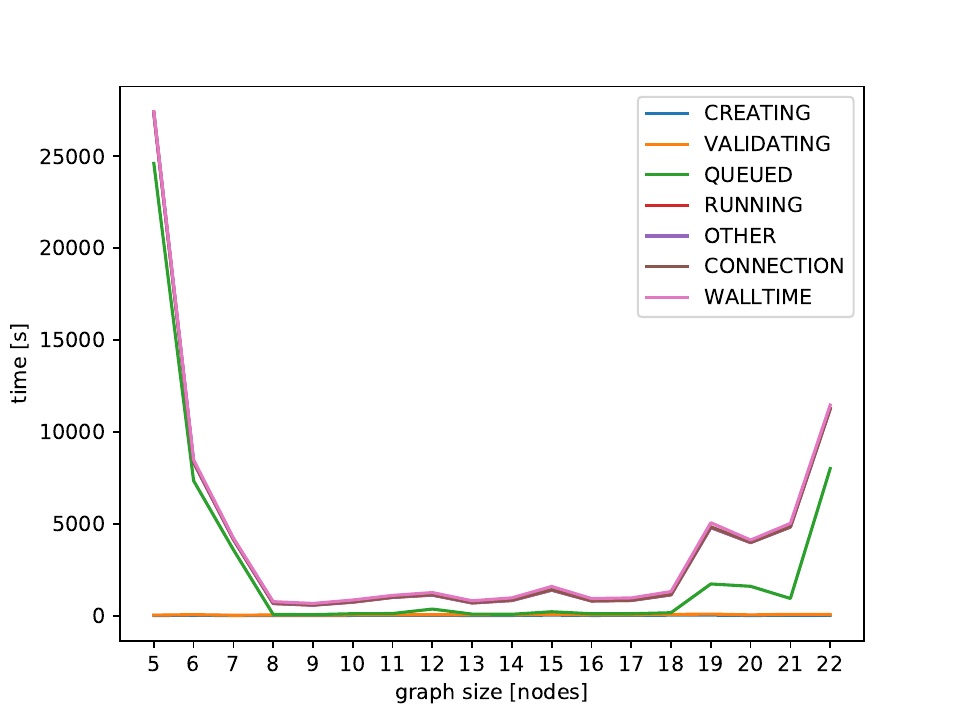}
    \caption{Results of the DSP benchmark up to 22 nodes on the IBM Montreal.}
    \label{fig:montreal_dsp}
  \end{figure}

\begin{figure}[h]
    \includegraphics[width=0.4\textwidth]{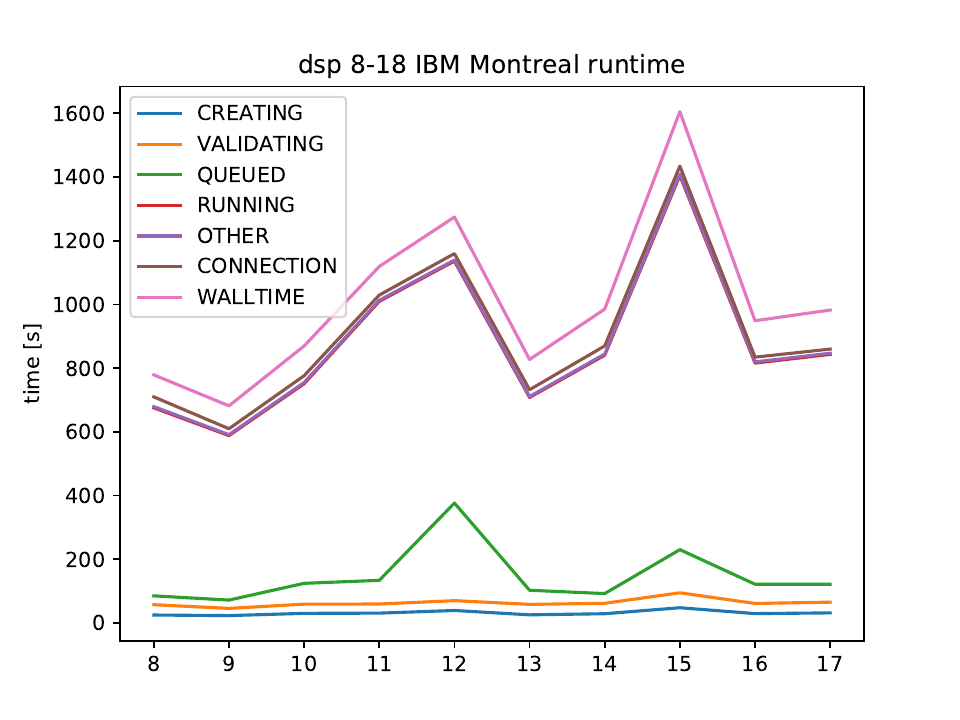}
    \caption{Results of the DSP benchmark 8 to 18 nodes on the IBM Montreal.}
    \label{fig:montreal_dsp_18}
  \end{figure}

The DSP results are presented for the IBM QASM simulator and the IBM Montreal. The results are not available for the IBM Nairobi hardware, as not enough qubits are present to run this benchmark. The runtimes for the DSP benchmark on the IBM QASM simulator are shown in Figure \ref{fig:QASM_dsp_15}. This figure shows large queue and run spikes at graph sizes 14 and 15. In order to show the results of graph sizes 5 to 13 in more detail, Figure \ref{fig:QASM_dsp_13} only shows the results for these sizes. The results for the IBM Montreal are presented in Figure \ref{fig:montreal_dsp}. This figure again shows large queue times for graph sizes 5 to 7 and 18 to 22. For this reason, Figure \ref{fig:montreal_dsp_18} only shows the results for graph sizes 8 to 17, for the same benchmark run.
With the implementation of the dominating set problem (DSP), we can see that the runtimes are significantly higher on the quantum hardware compared to the MCP benchmark. While the results for the IBM Montreal hardware and IBM QASM simulator in Figure~\ref{fig:QASM_dsp_15} and~\ref{fig:montreal_dsp} are mainly dominated by queue times, the cut-out results in Figure~\ref{fig:QASM_dsp_13} and~\ref{fig:montreal_dsp_18} clearly show larger runtimes compared to the MCP benchmark on the (simulated) hardware.  This result is to be expected, as the DSP QAOA circuit requires significantly more operations and qubits, as discussed before.
%QASM
  %\caption{Results of the DSP benchmark on the IBM QASM simulator}

%\begin{figure}[h]
%  \caption{Results of the DSP benchmark on the IBM Montreal}
%\end{figure}

\subsection{TSP}
\begin{figure}[h]
\includegraphics[width=0.9\linewidth]{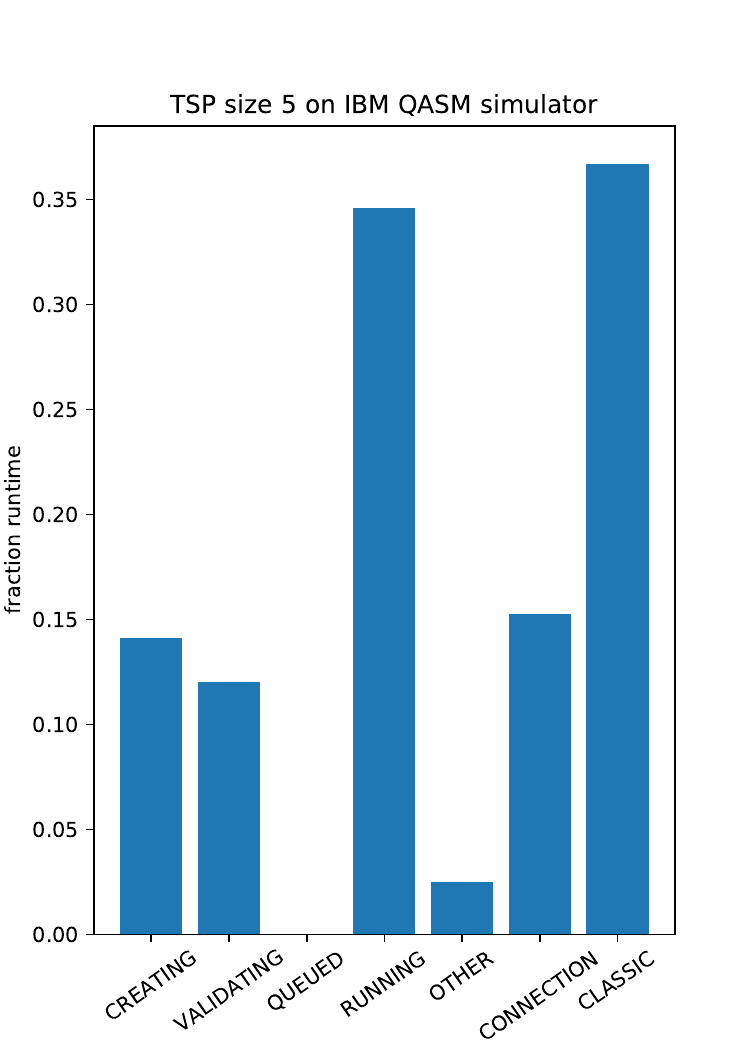}
\caption{Results of the TSP benchmark for 5 nodes on the IBM QASM simulator.}
\label{fig:QASM_sim_tsp}
\end{figure}

\begin{figure}[h]
\includegraphics[width=0.9\linewidth]{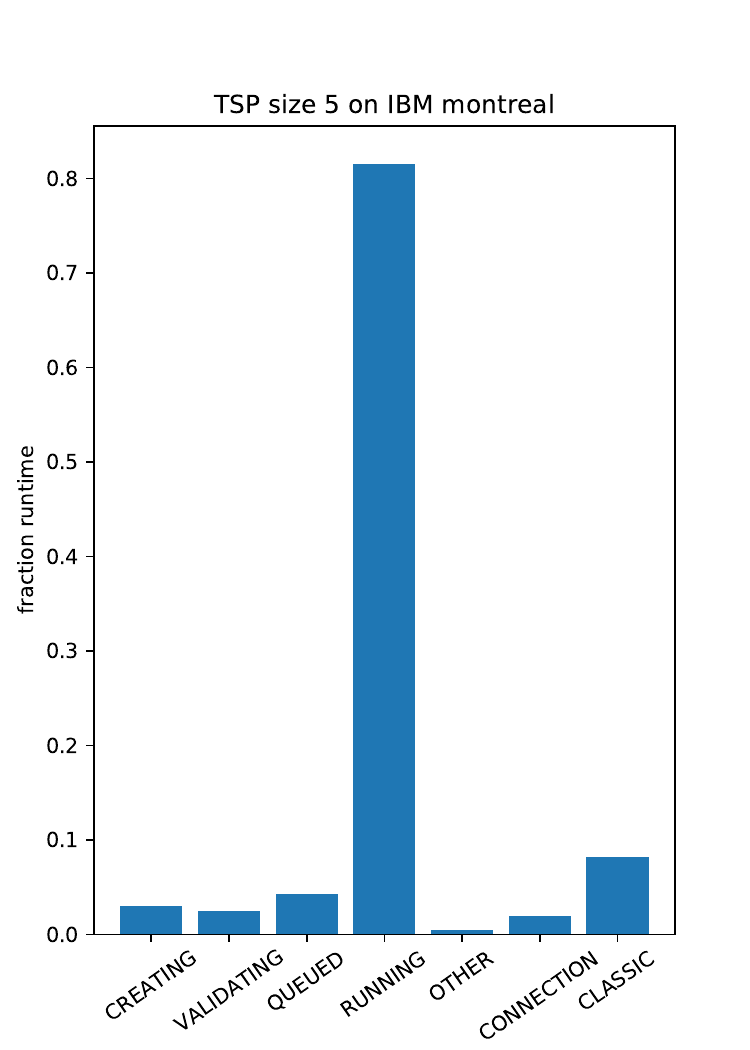}
\caption{Results of the TSP benchmark for 5 nodes on the IBM Montreal.}
\label{fig:tsp_montreal}
\end{figure}

As the traveling salesman problem (TSP) benchmark requires significantly more qubits, it could only be run on the IBM QASM simulator and the Montreal hardware for 5 nodes (25 qubits). The results are presented as fractions of the total runtime and are shown in Figure \ref{fig:QASM_sim_tsp} and \ref{fig:tsp_montreal}. This was done as only the results from graph size 5 were possible to measure. Therefore, the results cannot be presented as a line graph. The total runtime for the IBM QASM simulator was 21314.93s, while the total time spend on the IBM Montreal was 971.87s. Comparing these results, we see that on the hardware the main fraction of the computation (about 80\% on the Montreal hardware) is spent on the quantum hardware. It can also be seen that this fraction is significantly more than the MCP (which has a running time percentage of 67.39\% on the IBM Montreal) or DSP (which has a running time percentage of about 58.98\% on the IBM Montreal) benchmarks, as the TSP QAOA circuit is not only much larger in terms of qubits, but also in terms of operations. 

A notable difference between the runtime of the IBM QASM simulator and the IBM Montreal, is that significantly more time is spent on classical computations. This could indicate that the IBM QASM simulator takes proportionally less time to simulate the quantum circuit run compared to an actual hardware run.

\section{Outcome accuracy (score)}
This section presents the second part of our benchmark measurement called outcome accuracy (or score) achieved by the benchmark. The optimal scores are obtained by a brute force search. All measured solutions are obtained with parameter $p=1$.

\subsection{DSP}
%\begin{figure}[h]
%  \begin{subfigure}[b]{0.5\textwidth}
%    \includegraphics[width=\textwidth]{figures/score_QASM_dsp.pdf}
%    \caption{Score results for the IBM QASM simulator using the DSP benchmark.}
%    \label{fig:score_QASM_dsp}
%  \end{subfigure}
  %
%  \begin{subfigure}[b]{0.5\textwidth}
%    \includegraphics[width=\textwidth]{figures/score_montreal_dsp.pdf}
%    \caption{Score results for the IBM Montreal using the DSP benchmark.}
%    \label{fig:score_montreal_dsp}
%  \end{subfigure}
%  \caption{Comparison of the score results for DSP benchmark on the IBM QASM simulator and IBM Montreal}\label{dsp_score}
%\end{figure}

\begin{figure}[h]
    \centering
    \includegraphics[width=\linewidth]{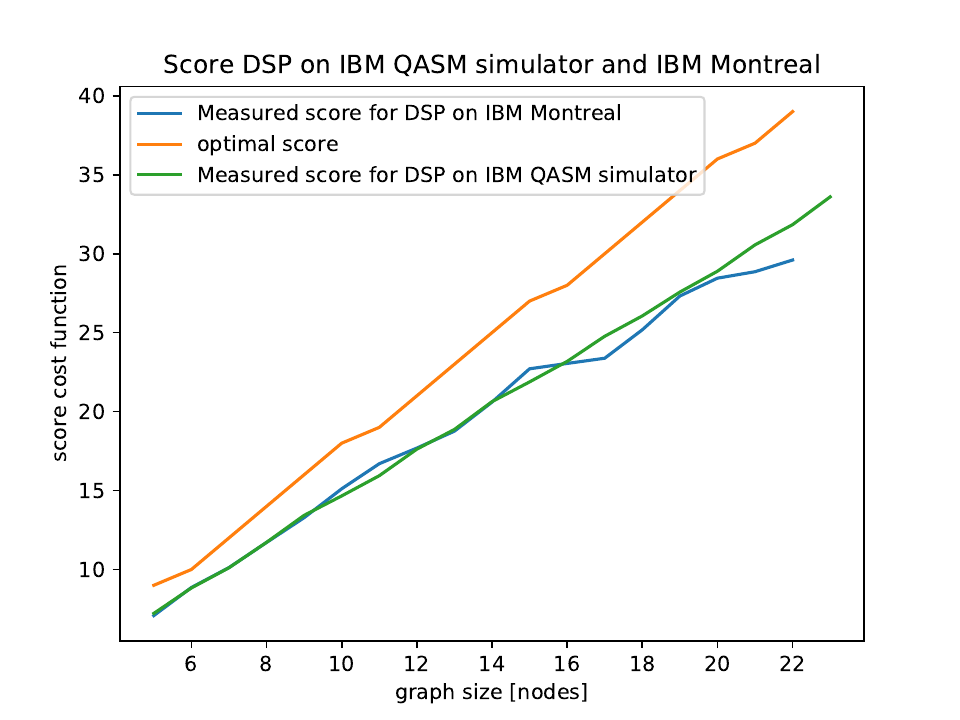}
    \caption{Comparison of the score results for DSP benchmark on the IBM QASM simulator and IBM Montreal}
    \label{fig:dsp_score}
\end{figure}

The results of the measured cost function score for the DSP benchmark on the IBM QASM simulator are shown in Figure \ref{fig:dsp_score}. The scores are calculated according to the cost evaluation for DSP presented in Section \ref{sec:app_measurements}. The scores are presented for graph sizes 5 to 23. The measured scores for the DSP benchmark on the IBM Montreal are also shown in Figure \ref{fig:dsp_score}. The measured scores for the Montreal hardware are shown for graph sizes 5 to 22. Both measured scores are compared to the optimal score. The score needs to be maximized, so the measured score is predictably lower. Interestingly, while the long runtimes in the previous section make it seem like DSP is a worse fit for QAOA, the algorithm appears to perform better in terms of score.

\begin{figure}[h]
    \centering
    \includegraphics[width=\linewidth]{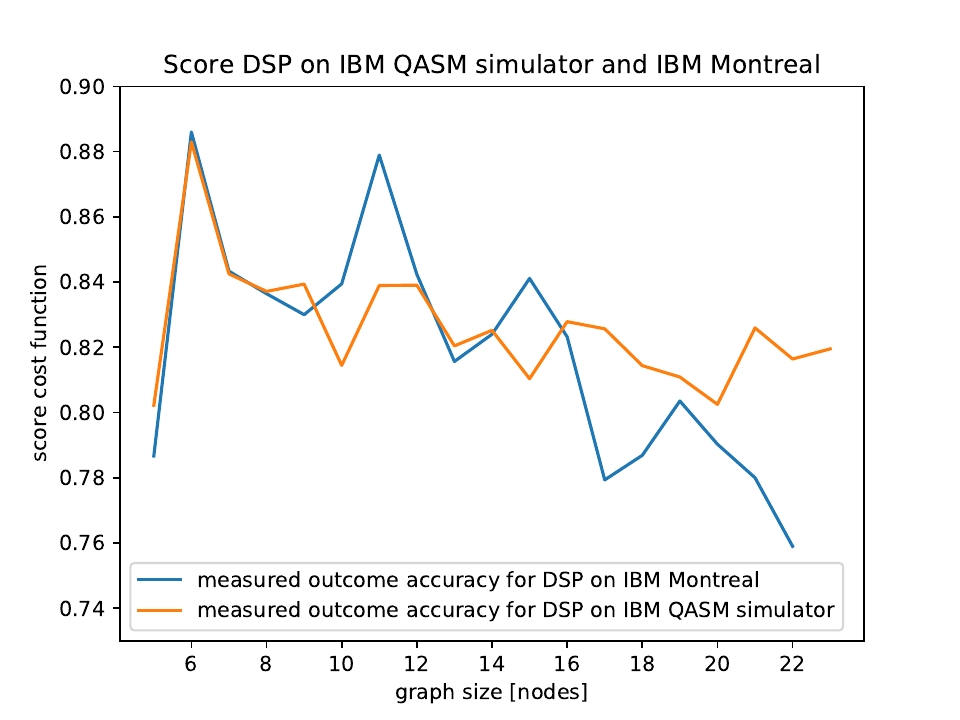}
    \caption{Comparison of the outcome accuracy results for DSP benchmark on the IBM QASM simulator and IBM Montreal}
    \label{fig:acc_dsp}
\end{figure}

The scores of Figure \ref{fig:dsp_score}, converted to the outcome accuracy, are presented in Figure \ref{fig:acc_dsp} for the IBM QASM simulator and the IBM Montreal. The outcome accuracy of the DSP benchmark results appear to be in a very similar order compared to the MCP benchmark results. The DSP benchmark results however, decay much less and prove to be more accurate for larger graph sizes compared to the MCP benchmark. The IBM QASM simulator retains its accuracy  for graph size 23 to around 82\% for the DSP benchmark, while it drops to 72.5 \% for the MCP benchmark. Similarly, on the IBM Montreal, the  accuracy for the DSP benchmark is measured at around 76\%, while the MCP benchmark results drop to around 58\%. It could be suggested to convert large MCP problems to DSP problems to maintain accuracy, but a large drawback is that the number of qubits and runtime significantly increase by doing so. For graphs with low connectivity per node, the DSP implementation might still be favored, as the increase of qubits is determined by the most connections to one node.

\subsection{TSP}
The TSP score results are again limited, due to the fact that the only graph size currently implementable is $5$ (25 qubits). The score found with the IBM QASM simulator for $p=1$ was averaged on 269.7, compared to an optimal score of 200. The score was found as follows: the fully connected 5-node graph consists of edges length 50. Any incorrect implementations of edges (i.e. edge $ij$ is connected but $ji$ is not) are punished by increasing the cost by 5 (and -5 if properly connected). Any edges that should not exist ($ii$, $jj$), are given length 100. Therefore the optimal score for size 5 becomes $5*50-10*5=200$. Comparing the score of 269.7 results in an error of $34.85\%$. This error is far from optimal, but the author of this algorithm claims that large $p$ is required for this algorithm to get acceptable results \cite{tsp_git}. Deciding on a proper value for parameter $p$ has no conclusive strategy, except that it is expected to grow linearly with the problem size. As such, the optimal value for $p$ is not further explored. As increasing $p$ increases the demand on the quantum hardware to support larger quantum circuit depth, it will be the challenge to the tester to increase $p$ as far as possible, up to the maximum supported depth of the hardware. This of course, should still be in proportion with the problem size.

%% file: additional_XACC_results.tex
\section{Additional XACC results}
\label{sec:app_xacc}

This section covers additional results to those presented in the main paper, implemented using the XACC library \cite{XACC}. This universal implementation shows how the QPack benchmark can be used to evaluate different quantum backends. Using the XACC library, the QPack benchmark was tested on 5 different simulators:
\begin{itemize}
    \item IBM QASM simulator (remote) \cite{qiskit}
    \item IonQ (remote) \cite{ionq_sim}
    \item aer (local) \cite{qiskit}
    \item qsim (local) \cite{qsim}
    \item qpp (local) \cite{gheorghiu2018quantum++}
\end{itemize}

Runtime results for above mentioned simulators are presented. The total runtime of the QAOA optimization is shown, as well as the average simulated quantum runtime for each optimizer iteration, as well as the number of iterations. All runtime results are presented on a logarithmic scale and simulators were set to execute 2048 shots. Similar to the main paper, the IBM QASM and the IonQ simulator were accessed remotely through the IBM Quatum experience and IonQ API respectively. The aer, qpp and qim simulators were all run locally as they were integrated in the XACC library.

As mentioned in the main paper, because of execution and measurements on different platforms and the results showing just a single QAOA run, a fair comparison between simulators cannot be made. Nevertheless, these results can give some insight on the performance of different backends and provide information on what simulator may be favorable to use in certain applications.\\

\begin{figure*}[h]
    \centering
    \includegraphics[width=0.7\textwidth]{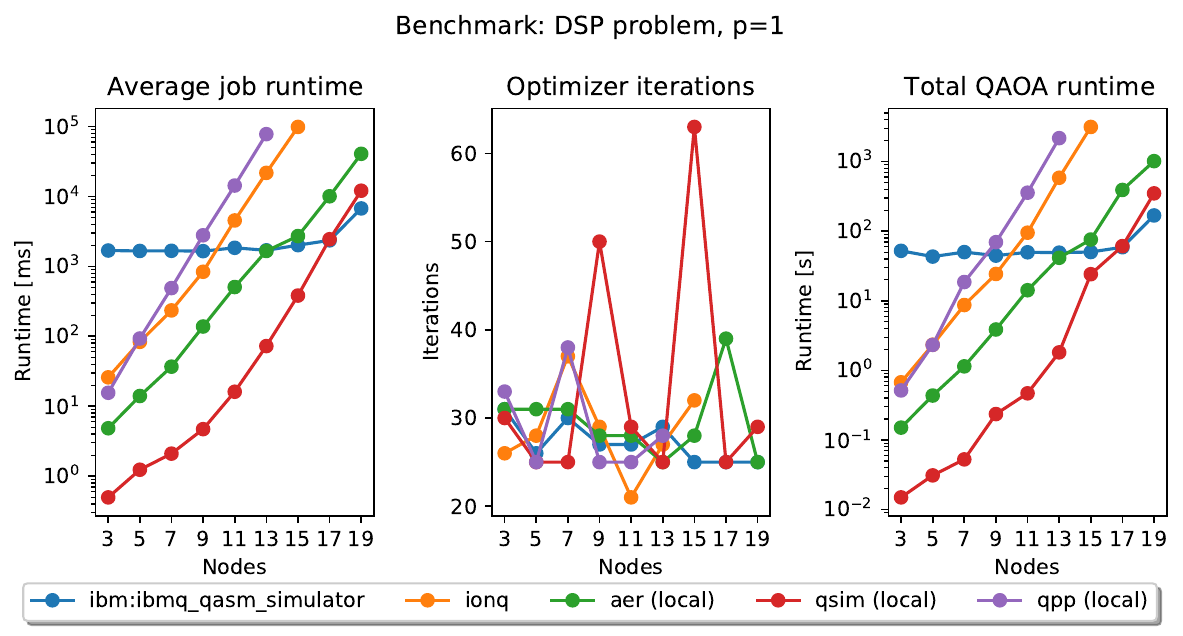}
    \caption{DSP benchmark runtime results for the QASM, IonQ, aer, qsim and qpp simulators for 3 to 19 nodes}
    \label{fig:xacc_dsp}
\end{figure*}

\begin{figure*}[h]
    \centering
    \includegraphics[width=0.7\textwidth]{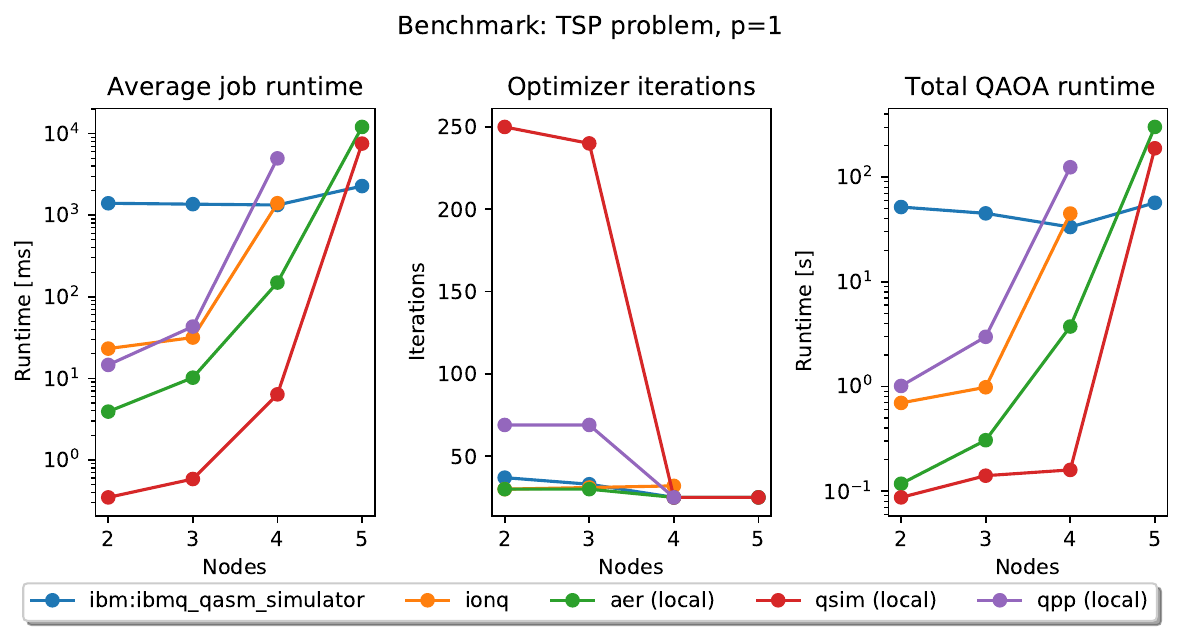}
    \caption{TSP benchmark runtime results for the QASM, IonQ, aer, qsim and qpp simulators for 2 to 5 nodes}
    \label{fig:xacc_tsp}
\end{figure*}

\subsection{DSP}
The results of the IBM QASM, IonQ, aer, qsim and qpp simulators for the DSP benchmark are shown in Figure \ref{fig:xacc_dsp}. These show the average simulated quantum runtime per iteration, the number of classical optimizer iterations and the total optimization time. Similarly to the MCP results (section \ref{sec:app_xacc}), the remote IBM QASM simulator shows a relatively logarithmic progression. In the DSP results, the qsim proves to be the fastest simulator for the tested sizes.\\

For the IBM QASM, aer and qsim simulators, sizes up to 19 nodes were possible. These larger sizes were not possible for the qpp and IonQ simulators, as these have shown difficulties for such large qubit requirements which limited them to 13 and 15 nodes respectively. With the implemented benchmark graphs, the quantum circuit would require $n+5$ qubits for $n$ nodes. The results interestingly show a much larger increase in runtime for the qsim simulator compared to the aer and IBM QASM simulator. Both the average quantum job runtime and the total runtime exceed the IBM QASM simulator runtimes. As this was something not seen in the MCP benchmark, this observation could mean that the qsim has much more difficulty simulating large circuit depths. This would be derived from the fact that the DSP benchmark has significantly larger circuit depth compared to the MCP benchmark for comparable numbers of qubits. Another reason could be that incidentally, the qsim simulator found more difficulty finding an optimal value for larger number of nodes, as the number of iterations at graph sizes 9 and 15 for the qsim are notably higher. The average runtime however, shows that even disregarding the number of iterations, the runtime is already increasing much faster compared to other simulators.

\subsection{TSP}

The results of the TSP benchmark for the IBM QASM, IonQ, aer, qsim and qpp simulators for 2 to 4 nodes are shown in Figure \ref{fig:xacc_tsp}. While these graph sizes appear to be trivial, the quantum requirement with the QAOA implementation are substantial. Interestingly, in these results, the runtime of the IonQ at 4 nodes (16 qubits) already exceeds the runtime of the IBM QASM simulator. This was not seen for the MCP benchmark. Most likely, the main contributing factor to this observation is the circuit length, similarly to earlier observations for the DSP benchmark.\\

The extended results for graph size 5 (25 qubits) are shown as well for the QASM, aer and qsim simulators. The IonQ and qpp results are not included for the same reasons in the extensions of the MCP and DSP benchmark. In these results, the IBM QASM simulator performs better at 5 nodes than both the aer and qsim simulator. The qsim simulator perform better than the ear simulator for the tested graph sizes, but  its runtime rises quicker. From the results of all previous benchmarks, the remote IBM QASM simulator performs better for qubits sizes, starting around 13 qubits. For smaller qubits the qsim benchmark performs better, but its runtime rises much quicker compared to other simulators. This can be observed more significantly when the benchmark involves larger circuit depths.